\documentclass[%
reprint,
floatfix,
showpacs,
amsmath,amssymb,
aps,
pra,
longbibliography
]{revtex4-1}

\usepackage{amsthm}
\usepackage{graphicx}
\usepackage{dcolumn}
\usepackage{bm}
\usepackage{bbold}
\usepackage{mathtools}
\begin{document}
	
	\title{Contact-based model for epidemic spreading on temporal networks}
	
	\author{Andreas Koher}
	\email{andreas.koher@tu-berlin.de}
	\affiliation{Institut f\"ur Theoretische Physik, Technische Universit\"at Berlin, Hardenbergstra{\ss}e 36, 10623 Berlin, Germany}%
	
	\author{Hartmut H. K. Lentz}%
	\affiliation{Institute of Epidemiology, Friedrich-Loeffler-Institut, S\"udufer 10, 17493 Greifswald - Insel Riems, Germany}%
	
	\author{James P.~Gleeson}
	\affiliation{MACSI, Department of Mathematics and Statistics, University of Limerick, Ireland}%
	
	\author{Philipp H\"ovel}
	\affiliation{School of Mathematical Sciences, University College Cork, Western Road, Cork T12 XF64, Ireland}
	\affiliation{Institut f\"ur Theoretische Physik, Technische Universit\"at Berlin, Hardenbergstra{\ss}e 36, 10623 Berlin, Germany}%
	\date{\today}
	
	\begin{abstract}
		We present a contact-based model to study the spreading of epidemics by means of extending the dynamic message passing approach to temporal networks. The shift in perspective from node- to edge-centric quantities enables accurate modelling of Markovian susceptible-infected-recovered outbreaks on time-varying trees, i.e., temporal networks with a loop-free underlying topology. On arbitrary graphs, the proposed contact-based model incorporates potential structural and temporal heterogeneities of the underlying contact network and improves analytic estimations with respect to the individual-based (node-centric)  approach at a low computational and conceptual cost. Within this new framework, we derive an analytical expression for the epidemic threshold on temporal networks and demonstrate the feasibility of this method on empirical data.
	\end{abstract}
	
	\maketitle
	
	
	
	\section{Introduction}
	\label{sec:intro}
	Accurate models of disease progression are valuable tools for public health institutions as they enable detection of outbreak origins \cite{BRO13, IAN17, HOR18, ALT14}, assessment of epidemic risk and vulnerability \cite{MAT18,ROG15, VAL15a} and, potentially, containment of the spreading at an early stage \cite{ALT14a, MAT18}. 
    Mitigation strategies can thus be evaluated and employed without the need to run a large number of Monte-Carlo (MC) realizations. 
    
    A fundamental challenge to mathematical epidemiologists is the accurate determination of the critical parameters that separate local and global epidemic outbreaks \cite{DIE90, PAS01, NEW02, CHA08a, MIL09a, VOL09, KAR14a}. To this end, the early Kermack-McKendrick model \cite{KER27} separates a population according to the disease status into compartments of susceptible, infected, and recovered individuals with mass-action equations to determine the transitions between them. A wide range of improvements has been proposed since, including the impact of stochasticity \cite{BAI75, SIM11, MIE09}, non-Markovian dynamics \cite{KIS15, SHE18a, KAR10a, GON11a, VAN13b} and, notably, heterogeneity in the contact structure \cite{MAY01, PAS01, KEE05, NEW06, KIS17a, POR16a}. 
    
    In recent years, the availability of mobility and contact data with high temporal resolution, so called \textit{temporal networks}, offers another opportunity to improve analytical predictions \cite{STO14, SEK16, HOL12, CAS12, ROC10, EAG09}. The timing of links between nodes matters, in particular when the network evolves on a similar time scale as the spreading dynamics, which led to an increasing interest in the interplay between disease and network dynamics \cite{DEL15a, GRO06b, MIL12b, KOH16, LEN16, DAR18}.
    
    One approach to model the states of a individual nodes in a network takes the corresponding probabilities directly as variables in a set of coupled dynamic equations \cite{WAN03b, MIE09, VAL15a, ROC15, CHA08a, GAN05, GOM10b, YOU11}. We will refer to this approach as the \textit{individual-based} (IB) model, though it is sometimes also called $N$-intertwined model \cite{MIE09} or quenched mean field \cite{GOM10a, PAS15}. However intuitive, the analytic predictability suffers from the simplifying assumption that epidemic states of adjacent nodes are independent. 
    
    Recently, a change from a node-centric to an edge-centric perspective has been discussed within different frameworks in order to overcome the inherent limitation of the IB model. These approaches include branching processes \cite{GLE14}, message passing \cite{KAR10a, LOK14}, belief propagation \cite{ALT14} and, the edge-based compartmental model \cite{MIL12b}. So far, however, edge-centric models are mostly limited to static topologies. It thus remains an open challenge to account simultaneously for topological and temporal properties of the underlying contact data and hence improve current predictions of the epidemic threshold \cite{BOK10, PRA10b, VAL15a, VAL15c, VAL18, SPE16}.

	In this paper, we generalize the dynamic message passing approach from \cite{LOK14} for discrete-time Markovian \textit{susceptible-infected-recovered} (SIR) spreading to time-evolving networks and derive the epidemic threshold within this new framework. The proposed model takes an edge-centric perspective, because the relevant dynamic equations are based on the set of edges. Furthermore, the framework integrates the complete temporal and topological information of the underlying network into the epidemic model. We will refer to our approach as \textit{contact-based} (CB) and compare numerical predictions with the widely used IB model that takes a node-centric perspective. Within the CB framework we then derive a new analytic expression of the epidemic threshold for temporal networks and show that the edge-centric approach improves existing results \cite{WAN03b, MIE09, BOK10, PRA10b, VAL15a, ROC15} at a low conceptual and numerical cost. Although both modelling frameworks can in principle account for contact weights that indicate the strength of a connection, we will focus on unweighted networks for simplicity and refer to the Appendix~\ref{sec:suppl_weighted_networks} for an extension of the model. The CB and IB models have been implemented in \textsc{Python} with the source code available on Github \cite{KOH18}.
	
	The remainder of this paper is structured as follows: First, we summarize the conceptual framework in Sec.~\ref{sec:concept} and formulate in Sec.~\ref{sec:Dynamic_Equations} the dynamic equations of the IB and CB models. Then, we derive the epidemic threshold for temporal networks within the CB framework in Sec.~\ref{sec:temporal_threshold_derivation}. We compare the edge- and node-centric approaches against Monte-Carlo (MC) simulations in Sec.~\ref{sec:application} and close with a discussion in Sec.~\ref{sec:discussion}. The appendix includes an extension to weighted contacts and heterogeneous epidemiological parameters in Sec.~\ref{sec:suppl_weighted_networks} as well as a network analysis of the German cattle trade data Sec.~\ref{sec:suppl_cattle_trade}. Further results and applications of the CB model are summarized in Sec.~\ref{sec:suppl_applications}.
	
	\section{Conceptual framework}
	\label{sec:concept}
	
	%
	We consider a temporal network $\mathcal{G} = [G(0), G(1), ..., G(T-1)]$ with $N$ nodes and $T$ snapshots sampled at a constant rate. A node $l \in \mathcal{N}$ represents an individual that is either susceptible, infected or, recovered at a given time $t$ with a corresponding probability $S_l(t)$, $I_l(t)$ and, $R_l(t)$, respectively. Emphasizing the important difference between temporal and static elements, we refer to \textit{contacts} as time-stamped links $(t,k,l) \in \mathcal{C} \subset \mathcal{T}\times\mathcal{N}\times\mathcal{N}$. Here, we denote with $\mathcal{N}$, $\mathcal{T}$ and, $\mathcal{C}$ the set of nodes, time stamps and, contacts, respectively. We further assume that every contact is of a constant duration and equal to the sampling time of the temporal network. With \textit{edges}, we refer to the corresponding static elements in the time-aggregated network. In other words, an edge $(k,l) \in \mathcal{E} \subset \mathcal{N}\times\mathcal{N}$ exists if and only if at least one (temporal) contact is recorded between $k$ and $l$. Here, we denote with $\mathcal{E}$ the set of edges. Moreover, we will assume directed edges throughout the paper and represent an undirected contact as two reciprocal contacts. Following the convention in \cite{KAR10a}, we denote with $k\rightarrow l$ a directed edge from $k$ to $l$, and we indicate edge-based quantities in a similar fashion.

    As the stochastic process, we assume a discrete-time SIR model. Here, a susceptible node that is in contact with an infected neighbor contracts the disease with a constant and uniform (per time-step) probability $\beta$. Furthermore, we treat the transmission events from multiple infected neighbors as independent and similarly, we interpret potential (integer) edge-weights as independent infection attempts (see Appendix~\ref{sec:suppl_weighted_networks}). Also, we do not account for secondary infections within one time step, i.e., only direct neighbors can be affected. Once infected, the individual recovers with a uniform and constant probability $\mu$ independently of the infection process and acquires henceforth a permanent immunity.
    
    Concerning the contact data, we will focus our numerical analysis first on a face-to-face interaction network between 100 conference participants \cite{ISE11}. This so-called proximity graph has a resolution of 20s and the observation time is limited to the first 24h. If necessary, we extend the data set with a periodic boundary condition in time. The time-resolved contacts enable the study of spreading of airborne diseases as well as the propagation of ideas and rumors. The data is available on \textsc{sociopatterns.org} and using the source code on \cite{KOH18}, results of this paper can be easily reproduced.
    
\begin{figure}[!ht]
    	\centering
    	\includegraphics[width=.8\columnwidth]{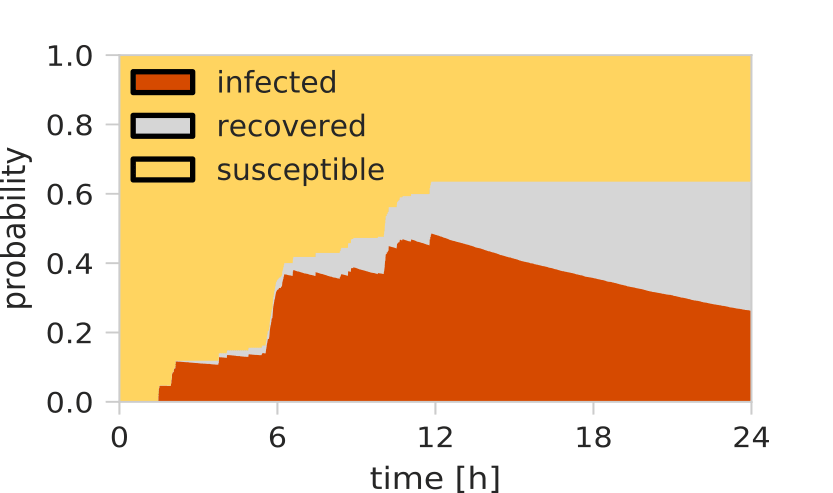}
    	\caption{Illustrative examples of a simulated epidemic outbreak from a single initially infected node. Colors give probability that another arbitrarily selected node is in the susceptible (yellow), infected (red) or recovered (grey) state, respectively. Simulation parameters: $\mu = 2.85 \cdot 10^{-4}$, $\beta = 100 \mu$, $10^4$ MC realizations} 
    	\label{fig:Fig1}
    \end{figure}

    As an illustrative example, we present in Fig.~\ref{fig:Fig1} the time-dependent probability that a selected node in the proximity graph is either susceptible (yellow), infected (red) or, recovered (gray). The results are derived from $10^4$ Monte-Carlo simulations with the same initially infected node. The trajectories reflect the bursty activity of the underlying temporal network \cite{ISE11} within the first 12h and the subsequent inactive night time.
    
    As a second source of data with direct relevance to public health, we consider an excerpt of the national German livestock database HI-Tier (\textsc{www.hi-tier.de}). This temporal network comprises the movement of cattle between farms in Germany for the year 2010 with daily resolution. Within the observation window of 365 days more than 3 million transactions have been recorded between over 180,000 farms and traders, respectively. For more details on the graph see Appendix~\ref{sec:suppl_cattle_trade}. Cattle trade is considered an important transmission route for livestock-related diseases such as foot-and-mouth disease (FMD), which broke out in Great Britain in 2001 with estimated costs of 8 billion British Pounds \cite{KEE08a}. Therefore, the analysis of the corresponding spatio-temporal graphs is highly relevant to public health institutions.

	\section{Dynamic equations}\label{sec:Dynamic_Equations}
	
	In this section, we will present the mathematical framework to model the stochastic SIR process as outlined in the introduction and Sec.~\ref{sec:concept}. Our main focus is the CB model but, in order to facilitate a direct comparison between the node and the edge-based approach, we will begin with a short overview of the IB model.
	
	\subsection{Individual-based model}\label{sec:IBM}

	In the IB model the marginal probabilities $S_l(t)$, $I_l(t)$ and, $R_l(t)$ for all $l \in \mathcal{N}$ enter directly a set of $3\times N$ coupled dynamic equations. The probability to transmit pathogens from node $k$ to $l$ upon a temporal contact is given by $\beta I_k(t)$. For convenience, we introduce an indicator function with $a_{k\rightarrow l}(t) = 1$ if a (directed) contact from $k$ to $l$ exists at time $t$ and $a_{k\rightarrow l}(t) = 0$ otherwise. Then, the probability for node $l$ to receive no infection at time $t$ from any of its neighbors factorizes to $\prod_k [ 1 - \beta a_{k\rightarrow l}(t) I_k(t)]$ and $k \in \mathcal{N}$. With this, the marginal probability $S_l(t+1)$ can be expressed by the probability $S_l(t)$ to be susceptible in the previous time step $t$ and not contract the infection within the interval $[t,t+1)$. In the IB model, the joint probability factorizes by assumption and we obtain
	\begin{equation}
	S_l(t+1) = S_l(t) \prod_{k \in \mathcal{N}} [ 1 - \beta a_{k\rightarrow l}(t) I_k(t)]. \label{eq:IBM1}
	\end{equation}
	Here, the crucial simplification is to treat the epidemic states of $l$ and its neighbors as mutually independent, which is sometimes referred as neglecting dynamic correlations \cite{CAI16a}.
	
	
	The marginal probability $I_l(t+1)$ follows from two independent contributions: (i) The out-flux $\mu I_l(t)$ indicates the transition from the infected to the recovered state. (ii) The in-flux $\Delta S_l(t) = S_l(t) - S_l(t+1)$ reflects the probability that node $l$ is newly infected at time $t+1$. Combining both contributions leads to
	\begin{equation}
	I_l(t+1) = (1-\mu) I_l(t) + S_l(t)\left\{ 1 - \prod_{k \in \mathcal{N}}[ 1 - \beta a_{k\rightarrow l}(t) I_k(t)] \right\}. \label{eq:IBM2}
	\end{equation}
	
	The set of $2 \times N$ coupled dynamic equations in  Eqs.~\eqref{eq:IBM1} and \eqref{eq:IBM2} thus constitutes the IB model for temporal networks. The remaining marginal probability $R_l(t)$ to find node $l$ in the recovered state follows from the conservation condition $S_l(t) + I_l(t) + R_l(t) = 1$ for all $l \in \mathcal{N}$. Finally, we will assign a probability $z_l = S_l(0)$ that node $l$ is initially susceptible as well as $I_l(0) = 1 - z_l$ and $R_l(0) = 0$ throughout the paper.
	
	Though intuitive and in many cases sufficient from a modelling perspective, the limits of the IB model are difficult to estimate due to the ad-hoc factorization of the joint probability in Eq.~\eqref{eq:IBM1}. Even for the simplest network with two nodes connected by an undirected static edge, the IB approach can deviate significantly from the expected outcome as illustrated in \cite{SHR15}. In their example recovery is neglected for simplicity and only the first node is infected initially with some probability $0 < z_1 \leq 1$. Counter-intuitively, the probabilities to find each node in the infected state converge to $I_1(\infty) = I_2(\infty) = 1$ according to the IB model, independent of the initial condition $z_1$. This is because integrating Eqs.~\eqref{eq:IBM1} and \eqref{eq:IBM2} admits a probability flux from the outbreak location to the adjacent node and back to its origin again. This mutual re-infection, coined \textit{echo chamber effect} in \cite{SHR15}, appears because we neglect the fact that the probability $I_2$, to find the second node in the infected state, is conditioned on the state of the first node and thus the factorization in Eq.~\eqref{eq:IBM1} is not justifiable.
	
	In an arbitrary network an initially infected node leads to a cascade of secondary infections within which all marginal probabilities are highly correlated. An accurate model excludes these previously infected nodes from those that can potentially contract the infection in the future. We will discuss in the next section how a shift from a node-centric to an edge-centric view can take into account some such dependencies.

	\subsection{\label{sec:temporal_derivation}Contact-based model}

	We begin with a slightly different approach to the marginal probability $S_l(t)$. First, we note that $l$ is susceptible at time $t$, if it was susceptible initially (with probability $S_l(0) = z_l$) and has not contracted the infection from any of its neighbors up to time $t$. We assign the probability $\Phi_l(t)$ to the latter statement. Thus without introducing any approximation at this stage, we can write 
	\begin{equation}
	S_l(t) = z_l \Phi_l(t). \label{eq:CBM_psi} 
	\end{equation}
	
	In order to determine $\Phi_l(t)$, we make the assumption that the underlying time-aggregated graph is a tree (ignoring directionality). Then, different branches originating in node $l$ are independent as long as $l$ remains susceptible and thus $\Phi_l(t)$ factorizes. However, if node $l$ contracts a disease from a neighbor $k$ with some probability and passes it on to another node $k'$ then the corresponding probabilities $I_k$ and $I_{k'}$ are clearly correlated. A simple solution that allows different branches to nonetheless be treated as independent is to prevent a probability flow through the root node in the first place. From a graph-theoretic perspective, this corresponds to the (virtual) removal of all out-directed contacts from the root node. This approach does not modify the dynamics of the node under consideration, because it can still contract the disease and once infected, the recovery process is independent of the topology. The idea reduces, however, considerably the amount of bookkeeping that would otherwise be necessary, if we accounted for the correlations directly. The singular node $l$ is said to be a cavity node or in the cavity state \cite{KAR10a, LOK14}, a concept closely related to the test-node assumption \cite{MIL12b} and the idea of cut-vertices \cite{KIS15a}. With this, we can factorize $\Psi_l(t)$ and thus obtain
	\begin{equation}
	S_l(t) = z_l \prod_{k \in \mathcal{N}_l} \theta_{k\rightarrow l}(t). \label{eq:CBM_psi_factorized} 
	\end{equation}
	Here, we introduced the probability $\theta_{k\rightarrow l}(t)$ that no disease has been transmitted from node $k$ to the cavity node $l$ up to time $t$.
	
	The change in perspective towards an edge-centric analysis introduces new auxiliary dynamic quantities such as $\theta_{k\rightarrow l}(t)$. These are defined on the set of edges $\mathcal{E}$ of the time-aggregated network and thus the number of dynamic variables scales with $L$, the number of edges.

	In order to obtain a system of dynamic equations, we focus on our first edge-centric variable $\theta_{k\rightarrow l}$. Initially, no disease was transmitted such that $\theta_{k\rightarrow l}(0) = 1$ for all edges $(k,l) \in \mathcal{E}$. Henceforth, the dynamic quantity reduces only (i) upon a temporal contact, indicated by $a_{k\rightarrow l}(t)$ and (ii) if the adjacent node $k$ is infected without having transmitted the disease earlier to the cavity node $l$ - we denote the corresponding probability with $I_{k\rightarrow l}(t)$. Hence, the out-flow of probability is given by $\beta a_{k\rightarrow l}(t) I_{k\rightarrow l}(t)$, leading to our first dynamic equation
	\begin{equation}
	\theta_{k\rightarrow l}(t+1) = \theta_{k\rightarrow l}(t) - \beta a_{k\rightarrow l}(t) I_{k\rightarrow l}(t). \label{eq:CBM0}    
	\end{equation}
	
	Next, the probability $I_{k\rightarrow l}(t)$ that node $k$ is infective at time $t$ and has not yet passed the disease to the cavity node $l$ evolves according to three contributions: (i) It decreases with the recovery probability $\mu$ and (ii) with the probability $\beta$ to infect its target node upon a temporal contact. These processes are independent and may contribute simultaneously with the joint probability $\beta \mu$. (iii) $I_{k\rightarrow l}(t)$ increases with the probability $\Delta S_{k\rightarrow l}(t) = S_{k\rightarrow l}(t) - S_{k\rightarrow l}(t+1)$ that $k$ is newly infected by at least one of its incident neighbors \textit{excluding} the cavity node $l$. In sum and with the initial condition $I_{k\rightarrow l}(0) = 1 - z_k$ these contributions lead to:
	\begin{equation}
	I_{k\rightarrow l}(t+1) =  (1-\mu)[1-\beta a_{k\rightarrow l}(t)] I_{k\rightarrow l}(t) + \Delta S_{k\rightarrow l}(t). \label{eq:CBM2}
	\end{equation}
	
	Finally, we consider the probability $S_{k\rightarrow l}(t)$ that node $k$, adjacent to the cavity node $l$, is susceptible. Since $k$ is not affected by the state of $l$, it stays susceptible if it does not contract the disease from any of its remaining, incident neighbors $j \in \mathcal{N}_k \setminus l$. It has been shown in \cite{LOK14a} that the corresponding probability $\Phi_{k\rightarrow l}(t) = \prod_{j\in \mathcal{N}_k \setminus l} \theta_{j\rightarrow k}(t)$ factorizes and thus similar to Eq.~\eqref{eq:CBM_psi}, we find $ S_{k\rightarrow l}(t) = z_l \Phi_{k\rightarrow l}(t) $ or equivalently
	\begin{equation}
	S_{k\rightarrow l}(t+1) =  z_k \prod_{j\in \mathcal{N}_k \setminus l} \theta_{j\rightarrow k}(t+1). \label{eq:CBM1}
	\end{equation}
	
	
	The disease progression in the CB framework is fully characterized by Eqs.~\eqref{eq:CBM0} and \eqref{eq:CBM2}, a set of $2L$ coupled equations. Equation \eqref{eq:CBM1} is introduced here for convenience only and can be substituted into Eq.~\eqref{eq:CBM2}. Next, we return to the node-centric quantities. To this end, we note that $S_l(t)$ has been already determined in Eq.~\eqref{eq:CBM_psi_factorized}.
	The remaining marginals $I_l$ and $R_l$ are equivalent to the IB model and given by the conservation condition Eq.~\eqref{eq:infected} and Eq.~\eqref{eq:recovered}, respectively. Hence, we arrive at the following node-centric equations:
	\begin{align}
	S_l(t+1) &= z_l \prod_{k\in \mathcal{N}_l} \theta_{k\rightarrow l}(t+1) \label{eq:susceptibility}\\
	I_l(t+1) &= 1 - S_l(t+1) - R_l(t+1) \label{eq:infected} \\
	R_l(t+1) &= R_l(t) + \mu I_l(t). \label{eq:recovered}
	\end{align}
	
	The CB model is exact for temporal networks, whose undirected time-aggregated version is a tree-graph and therefore loop-free. Thus, the change of perspective allows us to take full account of dynamic correlations on tree topologies. Most realistic networks, however, contain a large number of loops such as triangles in social graphs, where two friends are likely to have many more friends in common. Here the CB model nevertheless appears to be ``unreasonably effective" (cf. \cite{MEL11}) and improves predictions significantly with respect to the IB approach as we will see in Sec.~\ref{sec:application}. For further extensions to the model that include heterogeneous infection and recovery probabilities as well as weighted contacts see Appendix~\ref{sec:suppl_weighted_networks}.
	

	\section{\label{sec:temporal_threshold_derivation}Epidemic threshold}
	
	The condition defining the epidemic thresholds can be derived by examining small perturbations around the disease-free state. If such perturbations die out then any outbreak remains local, but if the perturbation grows then a global epidemic may occur. For that, we consider a linearization of the dynamic Eqs.~\eqref{eq:CBM0} - \eqref{eq:CBM1}, which will give rise to a criticality condition, determining the \textit{epidemic threshold}. We begin with the ansatz $\theta_{k\rightarrow l}(t) = 1 - \delta_{k\rightarrow l}(t)$ and $z_l = 1-\epsilon_l$, where $\delta_{k\rightarrow l}(t), \epsilon_l \ll 1$ are small perturbations around the disease-free state for all nodes $l$ and edges $(k,l)$. Thus,  Eq.~\eqref{eq:CBM0} becomes:
	\begin{equation}
	\delta_{k\rightarrow l}(t+1) = \delta_{k\rightarrow l}(t) + \beta a_{k\rightarrow l}(t) I_{k\rightarrow l}(t). \label{eq:threshold_delta}
	\end{equation}
	
	In Eq.~\eqref{eq:CBM1} we keep the linear terms of the Taylor expansion, which transforms the product into a corresponding sum:
	\begin{subequations}
		\begin{align}
		S_{k\rightarrow l}(t+1) &= (1-\epsilon_k) \prod_{j\in \mathcal{N}_k \setminus l}
		\left[ 1 - \delta_{j\rightarrow k}(t+1) \right]\\
		&\approx 1 - \epsilon_k - \sum_{j\in \mathcal{N}_k \setminus l}
		\delta_{j\rightarrow k}(t+1) \label{eq:threshold_S_step2} \\
		&= S_{k\rightarrow l}(t) +
		\beta \sum_{j\in \mathcal{N}_k \setminus l} a_{j\rightarrow k}(t) I_{j\rightarrow k}(t). \label{eq:threshold_Delta_phi_S}
		\end{align}
	\end{subequations}
	
	In Eq.~\eqref{eq:threshold_S_step2} we substituted the dynamic Eq.~\eqref{eq:threshold_delta} and identified $S_{k\rightarrow l}(t)$ in the next step. From the resulting Eq.~\eqref{eq:threshold_Delta_phi_S} we can read the linearized form of $\Delta S_{k\rightarrow l}$, which allows us to decouple the dynamic equations for $I_{k\rightarrow l}$:
	\begin{equation}
	\begin{aligned}
	I_{k\rightarrow l}(t+1) \approx& (1-\mu)[1-\beta a_{k\rightarrow l}(t)] I_{k\rightarrow l}(t) + \\
	& + \beta \sum_{j\in \mathcal{N}_k \setminus l} a_{j\rightarrow k}(t) I_{j\rightarrow k}(t).
	\end{aligned}\label{eq:linear}
	\end{equation}
	
	Next, we rewrite the remaining set of $L$ coupled dynamic equations in a compact, matrix-based formulation and therefore introduce the vectors $\bm{I}(t)$ and $\bm{a}(t)$ with elements $I_{k\rightarrow l}(t)$ and $a_{k\rightarrow l}(t)$, respectively. To this end, we also express the linear operation $\sum_{j\in \mathcal{N}_k \setminus l} a_{j\rightarrow k}(t)$ in Eq.~\eqref{eq:linear}, which acts on the elements $I_{k\rightarrow l}(t)$ of the state vector, through the temporal unweighted non-backtracking matrix $\bm{B}(t)$:
    \begin{equation} \label{eq:non_backtracking_matrix}
    B_{kl,jk'}(t) = 
    \begin{cases}
    1, & \parbox[t]{.25\textwidth}{if $k' = k$, $j \neq l$, $(t,j,k') \in \mathcal{C}$, and $(k,l) \in \mathcal{E}$}\\
    0, & \text{otherwise.}
    \end{cases}
    \end{equation}
    In other words, $B_{kl,jk'}(t) = 1$ if the contact $(t,j,k')$  at time $t$ is incident on the edge $(k, l)$ (implying $k'=k$), and additionally $j \neq l$. Otherwise we have $B_{kl,jk'}(t) = 0$. It is only the non-backtracking property $j \neq l$ that sets $\bm{B}$ apart from the adjacency matrix of the ordinary line-graph. For temporal networks a subtle distinction has to be made between the first and the second index of the $L \times L$ dimensional matrix $\bm{B}$: The first corresponds to an out-directed (\textit{static}) edge $(k,l) \in \mathcal{E}$ of the underlying aggregated network and can be interpreted as a potential contact in the future. The second, however, is an incident (\textit{temporal}) contact $(t,j,k') \in \mathcal{C}$ from node $j$ to $k'$ at time $t$.
    We also introduce the diagonal matrix $\text{diag}(\bm{1} - \beta \bm{a}(t))$, with diagonal elements given by the vector $\bm{1}-\beta\bm{a}(t)$. Here, we denote with $\bm{1}$ the vector of all ones.
    With these definitions, we rewrite Eq.~\eqref{eq:linear} to
	\begin{equation}
		\bm{I}(t+1) = \left[ (1-\mu)\, \text{diag}(\bm{1} - \beta \bm{a}(t)) + \beta \bm{B}(t) \right] \bm{I}(t). 
	\end{equation}
	The explicit solution to the state vector $\bm{I}(T)$ at final observation time $T$ is formally given by $\bm{I}(T) = \bm{P}(\beta,\mu) \bm{I}(0)$, where the so-called \textit{infection propagator} $\bm{P}$ \cite{VAL15c} is introduced for notational convenience:
	\begin{equation}
	\bm{P}(\beta,\mu) = \prod_{t=0}^{T-1} \left[ (1-\mu)\, \text{diag}(\bm{1} - \beta \bm{a}(t)) + \beta \bm{B}(t) \right]. \label{eq:threshold_propagator}
	\end{equation}

	In order to evaluate the asymptotic behavior, we assume a periodic boundary condition in time, i.e., $\bm{B}(t) = \bm{B}(t+T)$. This allows us to assess the vulnerability of the temporal network through the spectral radius of the propagator $\bm{P}$. In particular, we find that a SIR-type outbreak is asymptotically stable under small perturbations, i.e., remains confined to a small set of nodes, as long as the spectral radius satisfies $\rho[\bm{P}(\beta, \mu)] < 1$. Thus, the phase transition is given by the criticality condition
	\begin{equation}
	1 = \rho \left( \prod_{t=0}^{T-1} \left[ (1-\mu)\, \text{diag}(\bm{1} - \beta \bm{a}(t)) + \beta \bm{B}(t) \right] \right). \label{eq:threshold}
	\end{equation}
	Note that for irreducible and non-negative matrices the largest eigenvalue is simple and positive according to the Perron-Frobenius theorem \cite{MEY01}. Assuming $0 \le \beta, \mu < 1$, a sufficient condition for temporal networks is to restrict contacts to the \textit{giant strongly connected component} (GSCC) of the underlying time-aggregated graph. In  Sec.~\ref{sec:epidemic_threshold} we will fix the recovery probability $\mu$ and determine the critical infection probability $\beta_{\text{crit}}$ as the root of $f(\beta) = 1 - \rho[\bm{P}(\beta, \mu)]$ for different empirical networks.
	
	We conclude this section with a discussion on the static network limit. In the so-called quenched regime, the disease evolves on a much faster time scale than the dynamic topology and thus operates on an effectively static network with $\bm{B}(t) \equiv \bm{B}(0) \equiv \bm{B}$ and $\bm{a}(t) \equiv \bm{1}$ for all times $t$. As in the temporal analysis, we restrict the network to the GSCC so that the Perron-Frobenius theorem \cite{MEY01} applies. In this limit the dynamic equations~\eqref{eq:CBM0} - \eqref{eq:CBM1} reduce to the \textit{dynamic message passing} formulation in \cite{LOK14}. Moreover, Eq.~\eqref{eq:threshold_propagator} becomes now a product $\prod_{t=0}^{T-1} \bm{P}_{\text{fast}}(\beta,\mu) = [\bm{P}_{\text{fast}}(\beta, \mu)]^T$ of $T$ identical, single time step propagators
	\begin{equation}
	\bm{P}_{\text{fast}}(\beta,\mu) = (1-\mu)(1 - \beta) \mathbb{1} + \beta \bm{B}, \label{eq:threshold_propagator_static}
	\end{equation}
	where $\mathbb{1} = \text{diag}(\bm{1})$ denotes the identity matrix.
	
	The spectral radius in Eq.~\eqref{eq:threshold} factorizes to $\rho[\bm{P}_{\text{fast}}(\beta, \mu)^T] = \rho[\bm{P}_{\text{fast}}(\beta, \mu)]^T$, and it follows that the criticality condition Eq.~\eqref{eq:threshold} reduces to $\rho[\bm{P}_{\text{fast}}(\beta, \mu)] = 1$. Furthermore, we find from basic linear algebra that $\rho[\bm{P}_{\text{fast}}(\beta, \mu)] = (1-\mu) (1 - \beta) + \beta \rho(\bm{B})$ and hence we obtain the corresponding static threshold condition
	\begin{equation}
	\left(
	\frac{\beta}{\beta  + \mu - \beta \mu}
	\right)_{\text{crit, fast}} = \frac{1}{\rho(\bm{B})}. \label{eq:quenched}
	\end{equation}
	The criticality condition in Eq.~\eqref{eq:quenched} deviates from the continuous-time result in \cite{KAR10a, KAR14a}. In the derivation presented here, the term $ \beta \mu $ in Eq.~\eqref{eq:quenched} accounts for the simultaneous events when a node infects a neighbor and recovers within the same time step.
	
	In contrast to the quenched regime, one can also consider the so-called annealed limit. Then, parameters $\beta$ and $\mu$ are sufficiently small such that no more than one infection or recovery event can take place within the observation time.
	Therefore, we expand the infection propagator to the first order in $\beta$ and $\mu$ and obtain:
	\begin{equation}
	\bm{P}_{\text{slow}}(\beta,\mu) = (1-T\mu)\mathbb{1} - T \beta\, \text{diag}(\bm{\bar{a}}) + T\beta \bm{\bar{B}}.
	\end{equation}
	Here, $\bm{\bar{a}} = 1/T \sum_t \bm{a}(t)$ and $\bm{\bar{B}} = 1/T \sum_t \bm{B}(t)$ denote the corresponding time averaged quantities. It is insightful to evaluate simple bounds for the set of parameters $(\beta,\mu)_{\text{crit, slow}}$ that satisfies the threshold condition $\rho(\bm{P}_{\text{slow}})=1$ in the annealed limit. With $1/T \le \bm{\bar{a}} \le 1$ for all elements in $\bm{\bar{a}}$ we thus find:
	\begin{equation}
	\left( \frac{\beta}{\beta + \mu} \right)_{\text{crit, slow}}
	\le \frac{1}{\rho(\bm{\bar{B}})} \le
	\left( \frac{\beta}{\beta/T + \mu} \right)_{\text{crit, slow}}. \label{eq:annealed}
	\end{equation}
	
	Assuming the upper bound in Eq.~\eqref{eq:annealed} overestimates the outbreak risk and can be considered a conservative choice from an epidemiological perspective. This limit is realized for a temporal network where every edge appears exactly once within the observation time, hence $\bm{\bar{a}} = 1/T$. The lower bound in Eq.~\eqref{eq:annealed} is exact in case of a static network, thus $\bm{\bar{a}} = 1$, and corresponds to the continuous-time result in \cite{KAR14a}. However, this limit underestimates the outbreak risk and therefore we conclude with a note of caution when applying results from static network theory directly to time-varying topologies.
	
	
	\section{Application}\label{sec:application}
	
	
	A big advantage common to both the node-centric IB and edge-centric CB modeling framework is a significant reduction in computational complexity compared to MC simulation. The CB model requires iteration through all edges at every time step and thus the time complexity scales with $\mathcal{O}(L T)$. The IB formulation and a single MC realization require $\mathcal{O}(\bar{C} T)$, where $\bar{C}$ denotes the average number of active contacts, which can be significantly smaller than $L$. Stochastic MC simulations on the other hand require a large number of realizations in order to provide reliable statistics. The computational disadvantage of MC simulations becomes even more apparent when we consider a complex quantity such as the epidemic threshold, which requires multiple ensemble averages for different sets of epidemic parameters in order to fit the critical infection probability (see Sec.~\ref{sec:epidemic_threshold}). Equally important however, is the accuracy of our analytic approach. Therefore, we will compare in this section estimations from the IB and CB mean-field model with MC simulations using empirical data as introduced in Sec.~\ref{sec:concept}.

	
	\subsection{Numerical analysis of the mean-field dynamics \label{sec:Application_Individual_Nodes}}
	
	
	
\begin{figure}[!htb]
		\centering
		\includegraphics[width=\columnwidth]{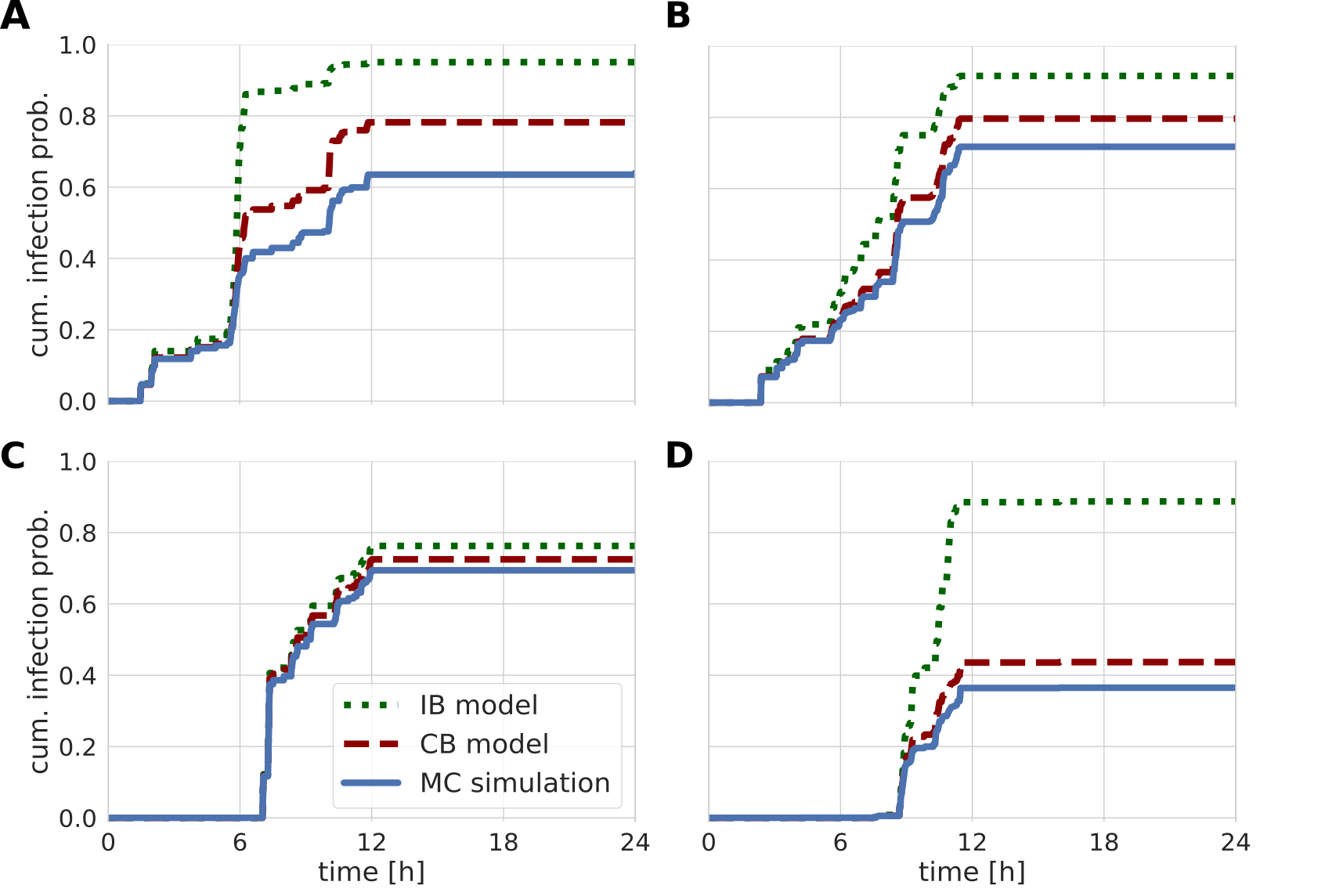}
		\caption{Epidemic trajectories for four exemplary individual nodes. We compare the cumulative infection probability from MC simulations (blue line) with estimations from the CB model (red dashed line) and the IB approach (green dotted line). Simulated results are averaged over $10^4$ MC realizations with the same outbreak location and disease parameters as in Fig.~\ref{fig:Fig1}.}
		\label{fig:Fig2}
	\end{figure}

	We begin with an analysis on the level of individual nodes. In Fig.~\ref{fig:Fig2}, we show the cumulative infection probability for a small number of example nodes from the conference data set given the same outbreak location. The selection is intended to present qualitatively different trajectories, also demonstrating that deviations between the two models vary considerably. The MC result (blue curve) in Fig.~\ref{fig:Fig2}A corresponds to the introductory example in Fig.~\ref{fig:Fig1}. Here, a comparison with the analytic estimation shows that the CB approach leads to a substantial improvement to the IB model. Also in Fig.~\ref{fig:Fig2}B-D, the trajectories are erratic, as they reflect the sudden changes in the underlying topology, highly individual and yet well approximated by the CB model. For all nodes in the network, we found that the CB model gives a closer upper bound to MC simulations because, unlike the IB framework, it accounts for dynamic correlations between nearest neighbor states.

	
\begin{figure}[!htb]
		\centering
		\includegraphics[width=\columnwidth]{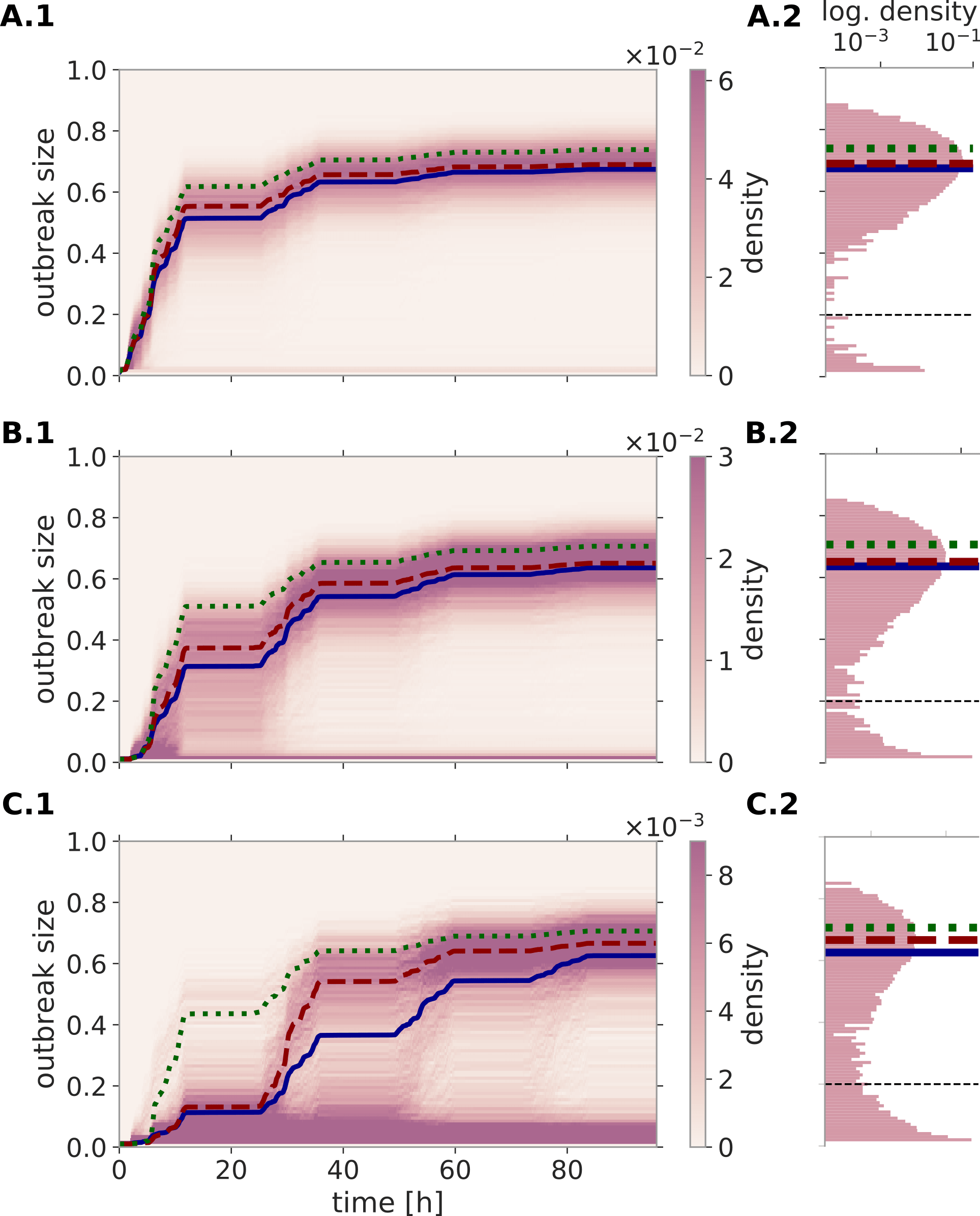}
		\caption{Left column: time-resolved and normalized outbreak size distributions for three different initially infected nodes. Epidemic parameters as in  Fig.~\ref{fig:Fig1} and $10^4$ MC realizations. The expectation value (blue line) assumes a cut-off at $20 \%$ of the population size (black dashed line). The CB and IB model are presented as a red dashed and green dotted line, respectively. Right column: Outbreak size distribution at the final observation time in logarithmic scale with corresponding expectation values.}
		\label{fig:Fig4}
	\end{figure}

	Dynamic mean-field models such as the IB and CB framework provide realistic expectation values only if stochastic fluctuations are negligible.
    In order to illustrate the limitations, we study epidemic outbreaks for three different initially infected nodes in Fig.~\ref{fig:Fig4} A, B, and C, respectively. The left column gives the time-resolved distribution of the outbreak size, and the right column presents the final distribution at the end of the three day observation period. For the ensemble average (blue line), we consider only realizations with more than 20 infected nodes overall. This threshold separates outbreaks that die out early due to stochastic fluctuations and thus permits a direct comparison with estimations from to the IB and CB framework in green and red, respectively.

	We chose the outbreak locations such that the degree of stochasticity increases from top to bottom. In Fig.~\ref{fig:Fig4}A.1, we find a narrow distribution around the ensemble average, which is well approximated by the mean-field models. Minor outbreaks due to early extinctions are well separated in Fig.~\ref{fig:Fig4}A.2 from large epidemics. In Fig.~\ref{fig:Fig4}B, the initially infected node leads to realizations with considerably stronger fluctuations and in Fig.~\ref{fig:Fig4}C it is barely possible to separate early extinctions at all. Additionally, we observe a second source of stochastic variation, namely the time at which a disease escalates and hence evolves into a global epidemic. As a consequence, early outbreak sizes may be overestimated significantly before the analytic trajectory approaches the expectation value again (see Fig.~\ref{fig:Fig4}C.1).
	
	Remarkably, the performance of both mean-field models varies significantly with the outbreak location, even for $R_0$ well above the epidemic threshold. At the late phase of an outbreak, however, the mean-field models provide good approximations and consistently with Fig.~\ref{fig:Fig2}, we find that the CB model outperforms the IB approach. In Appendix~\ref{sec:suppl_applications}, we demonstrate how a sufficiently large number of initially infected individuals improves significantly the predictability.
	
\begin{figure}[!htb]
		\centering
		\includegraphics[width=\columnwidth]{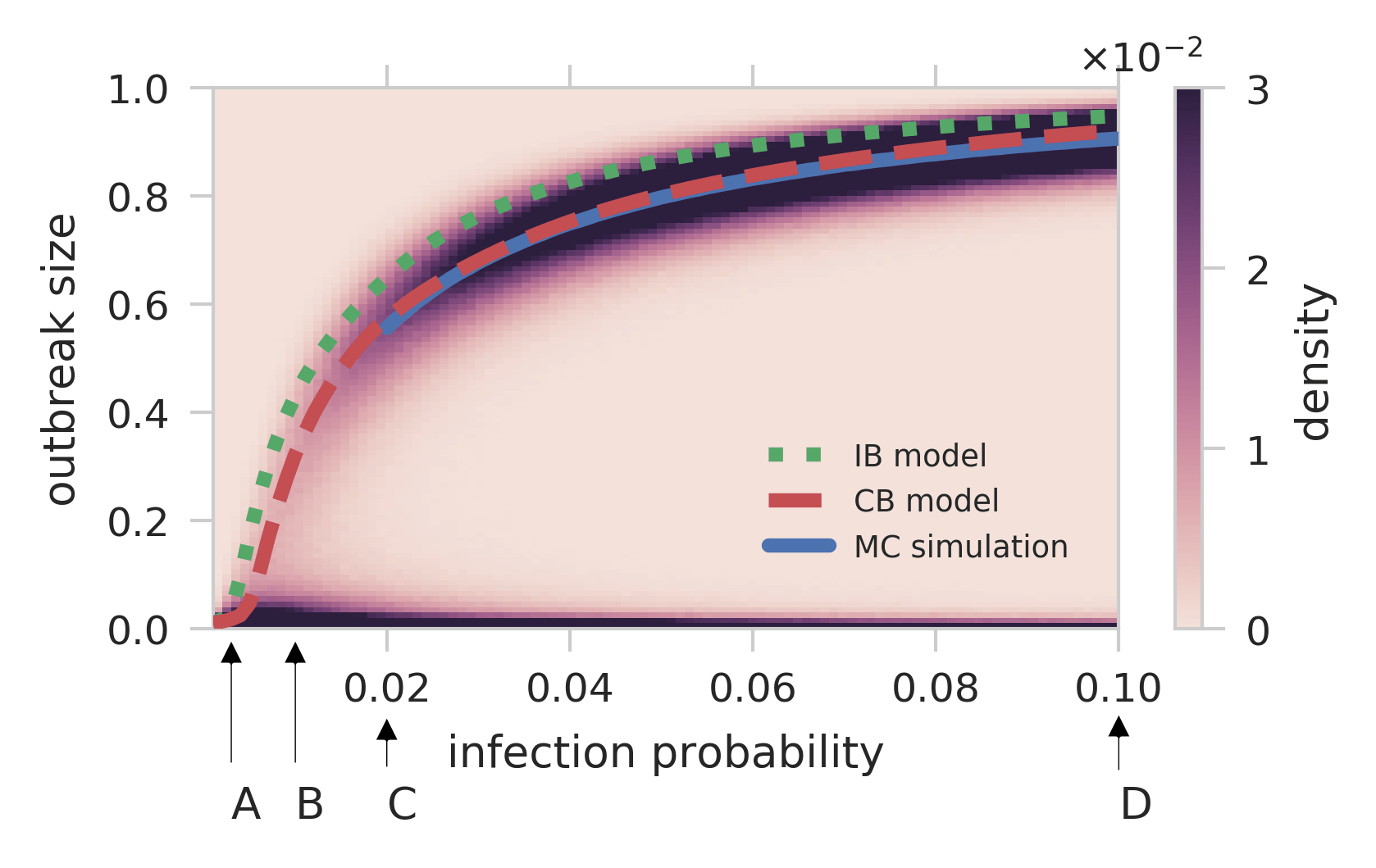}
		\caption{Distribution of final outbreak sizes as a function of the infection probability $\beta$. We perform $10^3$ MC realizations for every value of $\beta$, each starting with one randomly chosen outbreak location. For $\beta > 0.02$, we show the averaged final outbreak size (blue line) with a cut-off at $20$\% of the population size. 
		Estimations from the CB and IB model are presented as red dashed and green dotted lines, respectively.
		Labeled arrows at the bottom mark infection probabilities that correspond to Fig.~\ref{fig:attack_rates_slices_HT09} A,B,C and D, respectively.}
		\label{fig:attack_rates_HT09}
	\end{figure}

	Another source of stochasticity is the choice of disease parameters $\beta$ and $\mu$, respectively. For that, we focus on the final outbreak size, averaged over all outbreak locations. The distribution as a function of the infection probability $\beta$ (see Fig.~\ref{fig:attack_rates_HT09}) shows a percolation-like transition from localized spreading to epidemics that affect a considerable fraction of the network. We apply the same threshold as in Fig.~\ref{fig:Fig4} for a direct comparison between the averaged outbreak size and the mean-field models for $\beta > 0.02$. Here, we find that the difference between the expected size and the CB estimation is close to negligible, whereas the IB model consistently overestimates the expected value. 
	
\begin{figure}[!htb]
		\centering
		\includegraphics[width=\columnwidth]{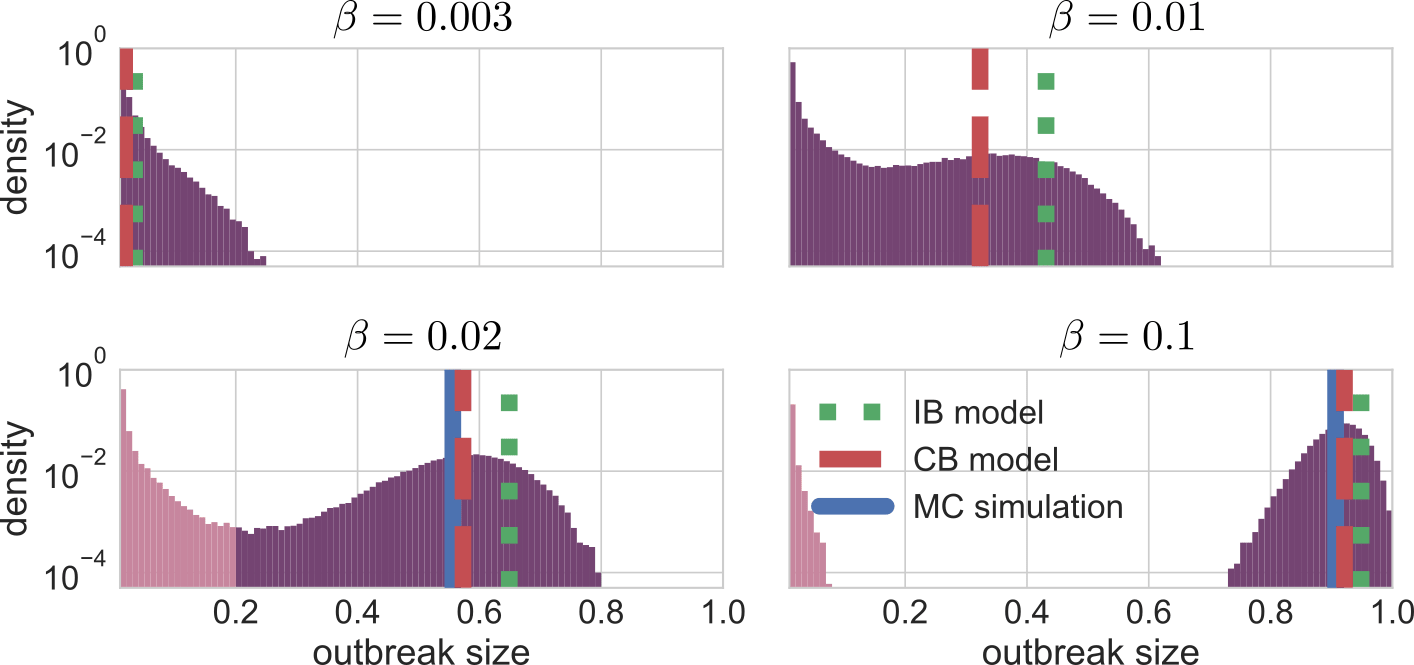}
		\caption{Distribution of final outbreak sizes for $\beta = 0.003$, $0.01$, $0.02$, and $0.1$, respectively. 
		For $\beta = 0.02$ and $0.1$ we mark part of the distribution with outbreak sizes below the given threshold of $20$\% by a lighter color tone and neglect this contribution to the averaged value (blue vertical line). The expected outbreak size from MC simulations and the estimations from the CB and IB model are plotted as blue, red dashed and green dotted vertical lines, respectively.
		}
		\label{fig:attack_rates_slices_HT09}
	\end{figure}

    A comparison at low values of the infection probability $\beta$ becomes unreliable as stochasticity impedes a reasonable distinction between minor and global outbreaks. In order to illustrate the effect, we present in Fig.~\ref{fig:attack_rates_slices_HT09} the outbreak size distribution for different values of $\beta$ as marked by the arrows in Fig.~\ref{fig:attack_rates_HT09}. 
	This representation highlights the transition from the sub- to the super-critical parameter domain: The unimodal distribution in Fig.~\ref{fig:attack_rates_slices_HT09}A characterizes localized outbreaks, whereas the bimodal distribution in Fig.~\ref{fig:attack_rates_slices_HT09}D clearly separates early extinctions and global epidemics. Next, we focus on the critical infection probability that marks the transition.
	
	\subsection{Epidemic threshold}
	\label{sec:epidemic_threshold}
	
\begin{figure}[!htb]
		\centering
		\includegraphics[width=\columnwidth]{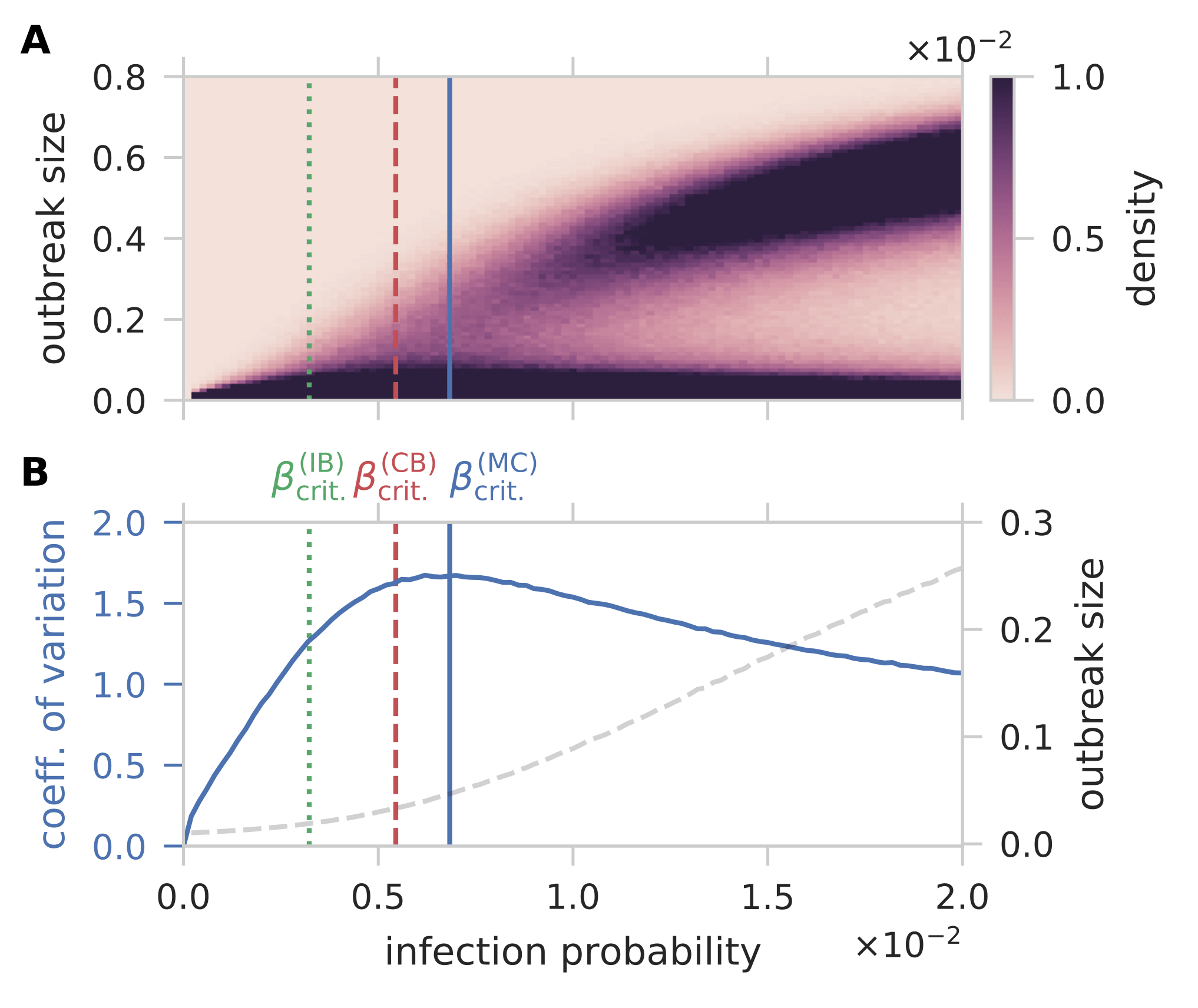}
		\caption{Estimation of the critical infection probability $\beta_{\text{crit}}$. \textbf{A}: Outbreak size distribution as in Fig.~\ref{fig:attack_rates_HT09} for small values of $\beta$. The vertical blue, red dashed and green dotted line mark the critical value according to MC simulations ($\beta^{\text{(MC)}}_{\text{crit.}}$), the CB model ($\beta^{\text{(CB)}}_{\text{crit.}}$) and IB approach ($\beta^{\text{(IB)}}_{\text{crit.}}$), respectively. \textbf{B}: From the distribution in A, we derive the coefficient of variation (blue line, left axis) and the mean outbreak size (grey dashed line, right axis).}
		\label{fig:susceptibility_HT09}
	\end{figure}

	In Fig.~\ref{fig:susceptibility_HT09}A, we present the region of  small $\beta$ from Fig.~\ref{fig:attack_rates_HT09}, in order to focus on the transition from localized outbreaks to the sudden emergence of global epidemics. We determine the critical infection probability $\beta_{\text{crit}}$ (vertical blue line) from the maximum of the relative standard deviation \cite{VAL15c}, also known as coefficient of variation (see blue line in Fig.~\ref{fig:susceptibility_HT09}B):
	\begin{equation}
	c_v = \frac{\sqrt{ \langle \sigma^2 \rangle - \langle \sigma \rangle^2}}{\langle \sigma \rangle} .\label{eq:chi}
	\end{equation}
	Here, we denote with $\langle \sigma \rangle$ and $\langle \sigma^2 \rangle$ the first and second moment of the outbreak size distribution. The coefficient of variation captures the intuition that fluctuations dominate the outbreak size distribution close to the transition. Indeed, $c_v$ diverges at the critical point for infinitely large networks, indicating a second-order phase transition \cite{STA94a}. 
	
	Analytically, we determine $\beta_{\text{crit}}$ from the spectral criterion in Eq.~\eqref{eq:threshold} for the CB model and similarly within the IB framework \cite{VAL15, VAL15c}. The comparison in Fig.~\ref{fig:susceptibility_HT09} shows that the IB and CB model, marked by a red dashed and green dotted line, respectively, underestimate the critical infection probability from MC simulations (blue line) and thus overestimate the outbreak risk.
	Consistent with our previous results, we can state that a shift from a node- to an edge-centric framework improves the analytic prediction. In Appendix~\ref{sec:suppl_applications}, we present similar results for different values of the recovery probability $\mu$. Next, we continue with a realistic application of the epidemic threshold to the German cattle trade network.
	

	\subsubsection{Application to German cattle trade \label{sec:cattle_trade}}
	
\begin{figure}[!htb]
		\centering
		\includegraphics[width=\columnwidth]{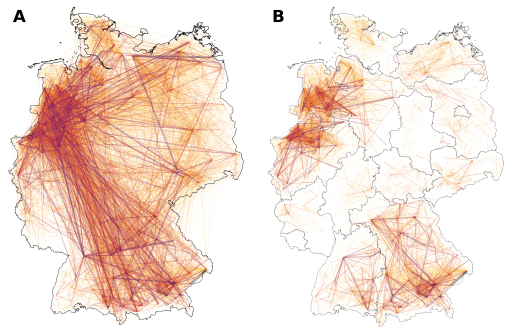}
		\caption{\textbf{A}: Cattle trade within Germany. Weighted edges correspond to directed trade relations within the year 2010, whereas the color indicates the accumulated number of traded animals. \textbf{B}: Cattle trade within the federal states of Germany. We confine the underlying time-aggregated graph to the GSCC and visualize here only edges with a flux of at least 50 animals.}
		\label{fig:German_network_edges}
	\end{figure}

	We now consider a completely different data set, where the system size is large and contacts are sparse over time. Our example is a cattle trade network, where the movements of animals between farms in Germany are recorded on a daily basis. Next, we isolate the trade within each federal state of Germany as visualized in Fig.~\ref{fig:German_network_edges} and restrict trade to the GSCC of the underlying aggregated graph. Disregarding the smallest networks (those with less than 27 nodes), we thus obtain 12 time-varying graphs with sizes varying from 254 to 27,863 nodes and highly heterogeneous topological and temporal features (see Appendix~\ref{sec:5_suppl_cattle_trade}).
	
	As in the previous section, we assume that premises can be either susceptible, infected or recovered and trade events facilitates the transmission of a disease. Unlike before however, we take into account the number of traded animals during each transaction, i.e., the weight $w_{k\rightarrow l}$ of a (temporal) contact from node $k$ to $l$. To this end, we modify the infection propagator in Eq.~\eqref{eq:threshold_propagator} and replace $\beta$ by $1-(1-\beta)^{w_{k\rightarrow l}}$ (see Appendix~\ref{sec:suppl_weighted_networks} for more information).
	In a potential outbreak, we assume that an infected node is detected with a constant probability $\mu$ each day after which it would be isolated and thus removed from the network. As a consequence highly infectious diseases such as FMD can be modelled as SIR-type epidemics \cite{KEE08a}.
	
\begin{figure}[!htb]
		\centering
		\includegraphics[width=\columnwidth]{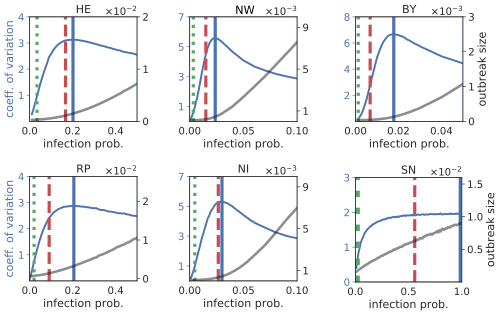}
		\caption{Detailed threshold analysis for six selected federal states with $\mu = 1/28$ (cf. Tab.~\ref{tab:bundeslaender}).  
		Simulated mean outbreak size (grey line, right axis) and coefficient of variation (blue line, left axis) averaged over all initially infected nodes. The critical infection probability from MC simulations, the IB and CB models are presented as vertical blue, green dotted and red dashed lines, respectively.}
		\label{fig:cattle_threshold_comparison}
	\end{figure}

	In Fig.~\ref{fig:cattle_threshold_comparison}, we compare the critical infection probability similar to Fig.~\ref{fig:susceptibility_HT09} for six selected federal states with different transition characteristics. The critical value derived from MC simulations varies between $\beta_{\text{crit.}} = 0.018$ (BY) and $\beta_{\text{crit.}} = 1.0$ (SN). The latter indicates that outbreaks remain localized for every choice of $\beta$ due to sparse intra-state trade.
	
\begin{figure}[!htb]
		\centering
		\includegraphics[width=\columnwidth]{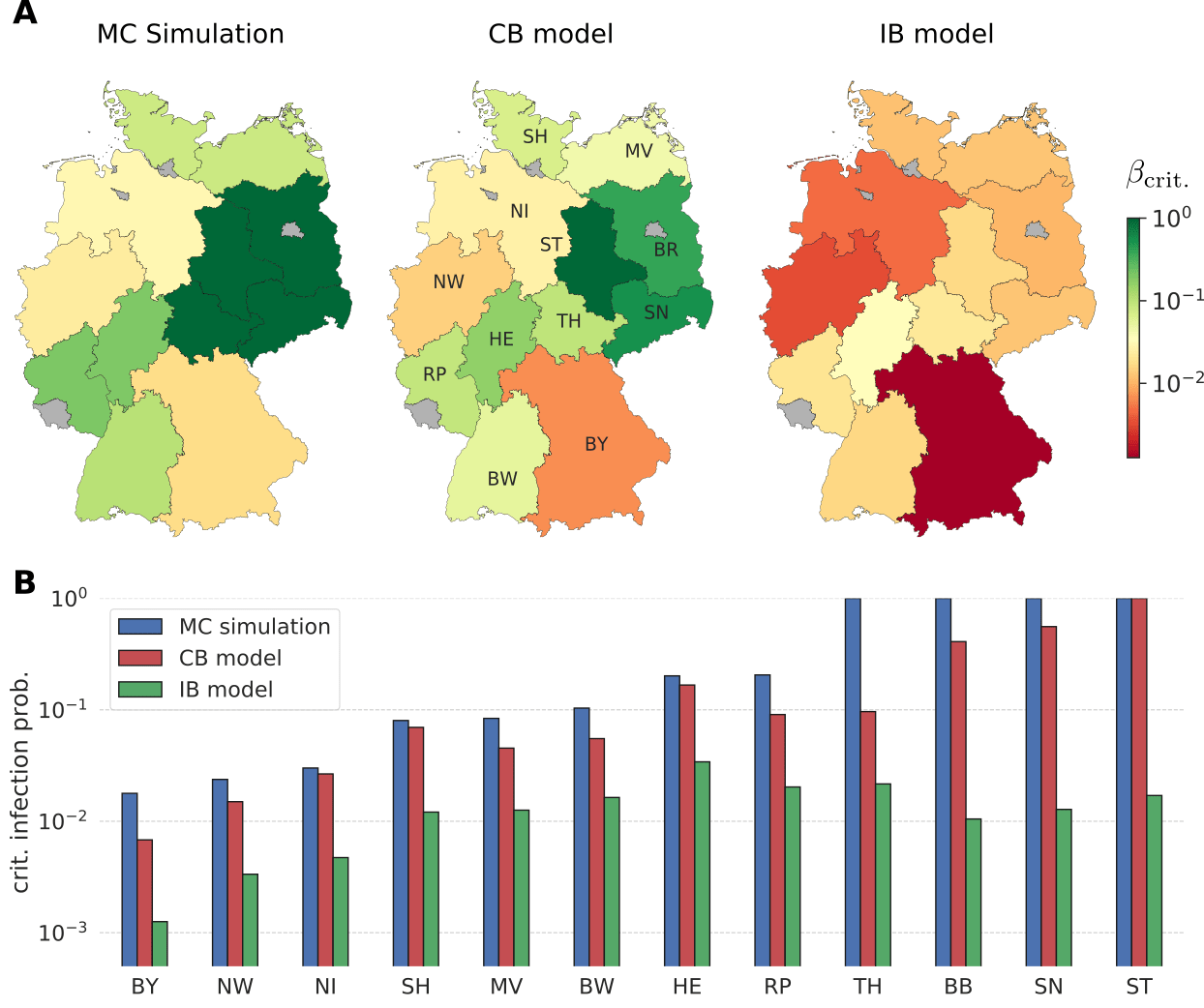}
		\caption{\textbf{A}: Spatial variation of epidemic risk due to cattle trade in Germany for $\mu = 1 / 120$. Federal states are colored according to the critical infection probability $\beta_{\text{crit}}$ as determined from MC simulations, the CB and IB model, respectively (see Fig.~\ref{fig:cattle_threshold_comparison} and appendix Fig.~\ref{fig:HIT_threshold_analysis_28days} for details). The city states Berlin (B), Hamburg (HH) and Bremen (HB) as well as Saarland (SL) are excluded due to the small network size (see appendix Tab.~\ref{tab:bundeslaender}). \textbf{B}: Critical infection probability $\beta_{\text{crit}}$ in logarithmic scale, sorted from high (left) to low risk (right). Results from MC simulations, the CB and IB model are presented as groups of blue, red, and green bars, respectively.
		Disclaimer: A realistic vulnerability analysis requires, for instance, heterogeneous recovery probabilities and complex counter measures (see Appendix~\ref{sec:suppl_cattle_trade} for details).}
		\label{fig:Deutschland}
	\end{figure}

	As a potential application to public health institutions, we present in Fig.~\ref{fig:Deutschland}A the spatial variation of the epidemic risk in terms of $\beta_{\text{crit}}$. The quantitive comparison in Fig.~\ref{fig:Deutschland}B demonstrates that spectral methods provide a lower bound with a varying degree of accuracy depending on the network details. Despite their heterogeneity in size and activity, we find for all networks that the CB model outperforms the IB approach. The detailed results for all states as well as a similar analysis for $\mu^{-1} = 120$ are available in Appendix~\ref{sec:suppl_applications}. 
	
	
	
	\section{Conclusion \label{sec:discussion}}
	
	In this paper we have presented the contact-based (CB) model for epidemic SIR spreading on temporal networks as a conceptually similar framework to the widely used individual-based (IB) approach. Derived from the message-passing framework \cite{KAR10a, LOK14} it inherits its accuracy on loop-free topologies and improves analytic estimations with respect to the IB approach for arbitrary time-evolving graphs. Moreover, the focus on edge-based quantities that are updated in discrete time steps allows a seamless integration of temporal interactions. Structually similar to the node-centric IB model, the proposed CB approach poses a low conceptual barrier and admits application on large graphs.
	
	Importantly, the accuracy of the CB model improves existing approximations of the epidemic threshold, which is a crucial risk measure for public health institutions. To this end, we have studied the largest eigenvalue of the infection propagator matrix, which determines the disease propagation in the low prevalence limit and takes into account the full temporal and topological information up to the observation time. The largest eigenvalue can be easily found through repeated matrix multiplications, i.e., the so-called power method. Without relying on extensive Monte-Carlo (MC) simulations and a subsequent parameter fit, the critical value can thus be estimated with efficient, vectorizable tools from linear algebra that are available for most high-level programming languages.
	
	In the application section, we have focused first on a social contact-graph that can be used to analyze the propagation of airborne diseases as well as the spread of information. Our comparison between MC simulations and analytic estimations from the CB and IB model followed a bottom-up approach: We looked at (i) epidemic trajectories of individual nodes, (ii) the outbreak size given the same initially infected node, and (iii) the outbreak size for a range of infection probabilities, averaged over all outbreak locations. In all cases, the CB model provides a closer upper bound to MC simulations than the widely used IB model. All results based on the conference data set can be reproduced using the \textsc{Python} code provided in \cite{KOH18}
	
	As a particularly important application, we then compared analytic estimations of the critical infection probability with extensive MC simulations. To this end, we included a case study of livestock trade within 12 federal states in Germany with highly heterogeneous characteristics in terms size, density and temporal activity. Consistently, we found that the CB model improves the previous lower bound of the IB framework at a low conceptual and computational cost.

	Many excellent results have already been derived within the IB framework for empirical networks and in the context of random graphs (see \cite{KIS17b} for a recent review) that can further improve the CB model. We therefore expect that the conceptual simplicity of the CB framework allows to integrate features such as non-Markovianity \cite{SHE18a}, stochastic effects \cite{VAL12}, and estimations of uncertainty \cite{WIL17a} that are important to realistic disease models on temporal networks. Also, first steps towards higher order models that go beyond the tree-graph assumption have been proposed in the context of percolation theory \cite{RAD16} and diffusive transport \cite{SCH14k, LAM18a} and we expect these improvement to also be applicable to the CB model.
	
	\begin{acknowledgements}
		AK and PH acknowledge the support by Deutsche Forschungsgemeinschaft (DFG) in the framework of Collaborative Research Center 910. AK acknowledges further support by German Academic Exchange Service (DAAD) via a short-term scholarship. JPG acknowledges the support by Science Foundation Ireland (grant numbers 16/IA/4470 and 16/RC/3918). 
	\end{acknowledgements}

    %

	\clearpage
	\appendix
	\section{Weighted networks \& heterogeneous infection and recovery probabi \label{sec:suppl_weighted_networks}}
	
	In order to improve the predictive power of a network model it is often required to take into account additional information. In the main article, we focused on the temporal dimension. However, another important piece of information is the weight of a contact. The interpretation of weight can range from passenger numbers in the global air traffic network to the impedance in a network of electric components. The distribution of weights in static as well as temporal empirical networks, shows often a broad tail \cite{BAR04b} and as such the averaged edge weight can become meaningless due to large fluctuations. It is therefore often required to account for heterogeneous edge weights explicitly in epidemiological models. However, depending on the interpretation, weights may enter the model in different ways. Typically, a time-dependent edge weight $w_{k\rightarrow l}(t)$ is considered similar to the conductivity between two nodes $k$ and $l$ in an electric circuit. Translated to an epidemiological context, we would thus scale the infection probability linearly, i.e., $\beta a_{k\rightarrow l}(t)$ becomes  $\beta  w_{k\rightarrow l}(t)$ in a weighted network. Another approach, popular in the context of random walks and disease spreading, is to interpret an integer-valued weight $w_{k\rightarrow l}(t)$ as a number of parallel and unweighted edges that connect $k$ with $l$ \cite{GOM10b, VAL15a}. From an epidemiological viewpoint, this idea would translate to  $w_{k\rightarrow l}$ independent attempts to transmit the disease  at time $t$. Here, the infection probability $\beta a_{k\rightarrow l}(t)$ becomes $1 - (1 - \beta)^{w_{k\rightarrow l}(t)}$ in the weighted case. In the main article we apply the latter interpretation to calculate the epidemic risk in the context of livestock trade (see Fig.~\ref{fig:Deutschland}). Here, weights correspond to the number of animals traded, each of which can infect the target population independently. For small probabilities $\beta \ll 1$ the adjusted infection probability simplifies to $1 - (1 - \beta)^{w_{k\rightarrow l}(t)} \approx \beta w_{k\rightarrow l}(t)$.
	
	A second source of heterogeneity that is commonly considered includes heterogeneous infection and recovery probabilities, denoted as $\beta_{k\rightarrow l}$ and $\mu_k$, respectively. With these modifications the dynamic equations Eq.~\eqref{eq:CBM0}-\eqref{eq:CBM1} from the main text translate to
	\begin{subequations}
	\begin{align}
	    \theta_{k\rightarrow l}(t+1) &= \theta_{k\rightarrow l}(t) - \Psi_{k\rightarrow l}(t) I_{k\rightarrow l}(t) \label{eq:suppl_weighted_theta}\\
	    S_{k\rightarrow l}(t+1) &=  z_k \prod_{j\in \mathcal{N}_k \setminus l} \theta_{j\rightarrow k}(t+1)\\
	    I_{k\rightarrow l}(t+1) &=  (1-\mu_k)[1-\Psi_{k\rightarrow l}(t)] I_{k\rightarrow l}(t) + \Delta S_{k\rightarrow l}(t). \label{eq:suppl_weighted_I}
	\end{align}
	\end{subequations}
	
	Here, $\Psi_{k\rightarrow l}(t)$ denotes the probability that $k$ infects $l$ at time $t$ given that the former is infected and has not yet transmitted the disease. For weighted networks we can choose $\Psi_{k\rightarrow l}(t) = \beta_{k\rightarrow l} w_{k\rightarrow l}(t)$ or $\Psi_{k\rightarrow l}(t) = 1 - (1 - \beta_{k\rightarrow l})^{w_{k\rightarrow l(t)}}$ as discussed above. The linearization of Eq.~\eqref{eq:suppl_weighted_theta}-\eqref{eq:suppl_weighted_I} around the disease-free state leads to 
	\begin{equation}
	    \bm{I}(t+1) = \text{diag}[(\bm{1}-\bm{\mu})\circ(\bm{1} - \bm{\Psi}(t))] \bm{I}(t) + \bm{B}_{\beta}(t) \bm{I}(t).
	\end{equation}
	Here, the circle denotes the elementwise product. Moreover, the $L$-dimensional vectors $\bm{\mu}$ and $\bm{\Psi}(t)$ integrate the node- and edge-dependent values $\mu_k$ and $\Psi_{k\rightarrow l}(t)$, respectively. We also generalize the temporal non-backtracking matrix $\bm{B}_{\beta}(t)$ from Eq.~\eqref{eq:non_backtracking_matrix} to the weighted:
    \begin{equation}
    \left[ \bm{B}_{\beta}(t) \right]_{kl,jk'} = 
    \begin{cases}
    \Psi_{j\rightarrow k'}(t), \text{ if $k' = k$, $j \neq l$ and $(k,l) \in \mathcal{E}$}\\
    0, \text{ otherwise.}
    \end{cases}
    \end{equation}
	Note that the additional constraint $(t,j,k') \in \mathcal{C}$ from definition in Eq.~\eqref{eq:non_backtracking_matrix} enters indirectly through $ \Psi_{j\rightarrow k'}(t)$. 
	
	The largest eigenvalue $\rho$ of the infection propagator determines the asymptotic stability for small perturbations around the disease-free state. Accounting for heterogeneity in $\beta$, $\mu$ and contact-weights, the criticality condition Eq.~\eqref{eq:threshold} from the main text reads: 
	\begin{equation}
	1 = \rho \left( \prod_{t=0}^{T-1} \left\{
	\text{diag}[(\bm{1}-\bm{\mu})\circ(\bm{1} - \bm{\Psi}(t))] + \bm{B}_{\beta}(t) \right\} \right). \label{eq:suppl_weighted_threshold}
	\end{equation}
	
	Assuming $\beta \equiv \beta_{k\rightarrow l}$ for all edges $k \rightarrow l$, we can determine from Eq.~\eqref{eq:suppl_weighted_threshold} the critical (homogeneous) infection probability $\beta_{\text{crit}}$ given a weighted, temporal network with heterogeneous recovery probabilities $\mu_k$. Similarly, one can assume $\mu \equiv \mu_{k}$ for all nodes $k$ and thus derive the critical (homogeneous) value $\mu_{\text{crit}}$ with heterogeneity in the infection probability $\beta_{k\rightarrow l}$.
	
	\section{German cattle trade network}\label{sec:suppl_cattle_trade}
	
\begin{figure}[!htb]
		\centering
		\includegraphics[width=\columnwidth]{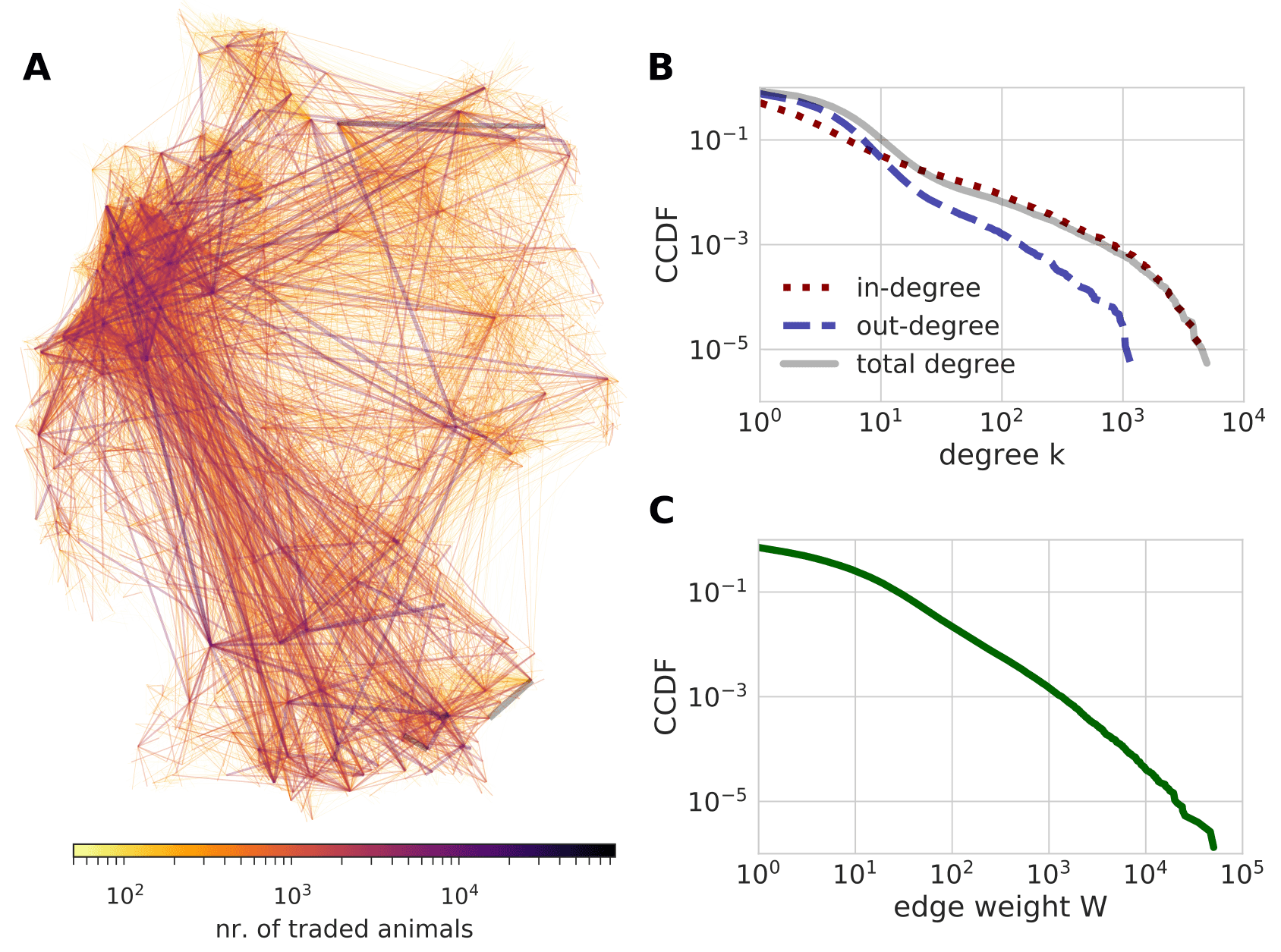}
		\caption{Degree and weight analysis of cattle trade in the year 2919. Trade data is aggregated over time, i.e., the year 2010. \textbf{A}: Visualization of edges using geo-location with at least 50 animals traded along each link. The color and edge width indicates the aggregated edge weight. \textbf{B}: Complementary cumulative distribution function (CCDF) of in-, out- and total degree. \textbf{C}: CCDF of the edge weight distribution.}
		\label{fig:network_degree}
	\end{figure}

	The system of traceability of cattle in the EU requires that each animal is identified with ear-tags and that each movement, birth or death event has to be reported within 7 days of the event to the national livestock database. We consider an excerpt of the national German livestock database HI-Tier (\textsc{www.hi-tier.de}) for the year 2010. The database is administered by the Bavarian State Ministry for Agriculture and Forestry on behalf of the German Federal States. It records $3.2$ million animal movements with a total of $13.4$ million traded animals between $183,454$ premises, such as farms, pastures, slaughter houses and traders within the observation window. Location of each animal holding was provided at the resolution of the municipality. 
	We consider each trade event between two premises a temporal contact and we identify an edge if at least one contact has been recorded. The distribution of edges is highly heterogeneous in terms of geography, degree and weight. In Fig.~\ref{fig:network_degree}A we observe clusters of trade activity mostly within and between North Rhine-Westphalia (NW), Lower Saxony (NI), Baden-W\"urttemberg (BW) and Bavaria (BY). The number of trading partners, i.e., the node degree, is broadly distributed as demonstrated in Fig.~\ref{fig:network_degree}B. Here, we differentiate between in-, out- and total degree. Similarly we find a broad distribution of edge weights in Fig.~\ref{fig:network_degree}C, i.e., the number of traded animals along a given edge. 
	
\begin{figure}[!htb]
		\centering
		\includegraphics[width=\columnwidth]{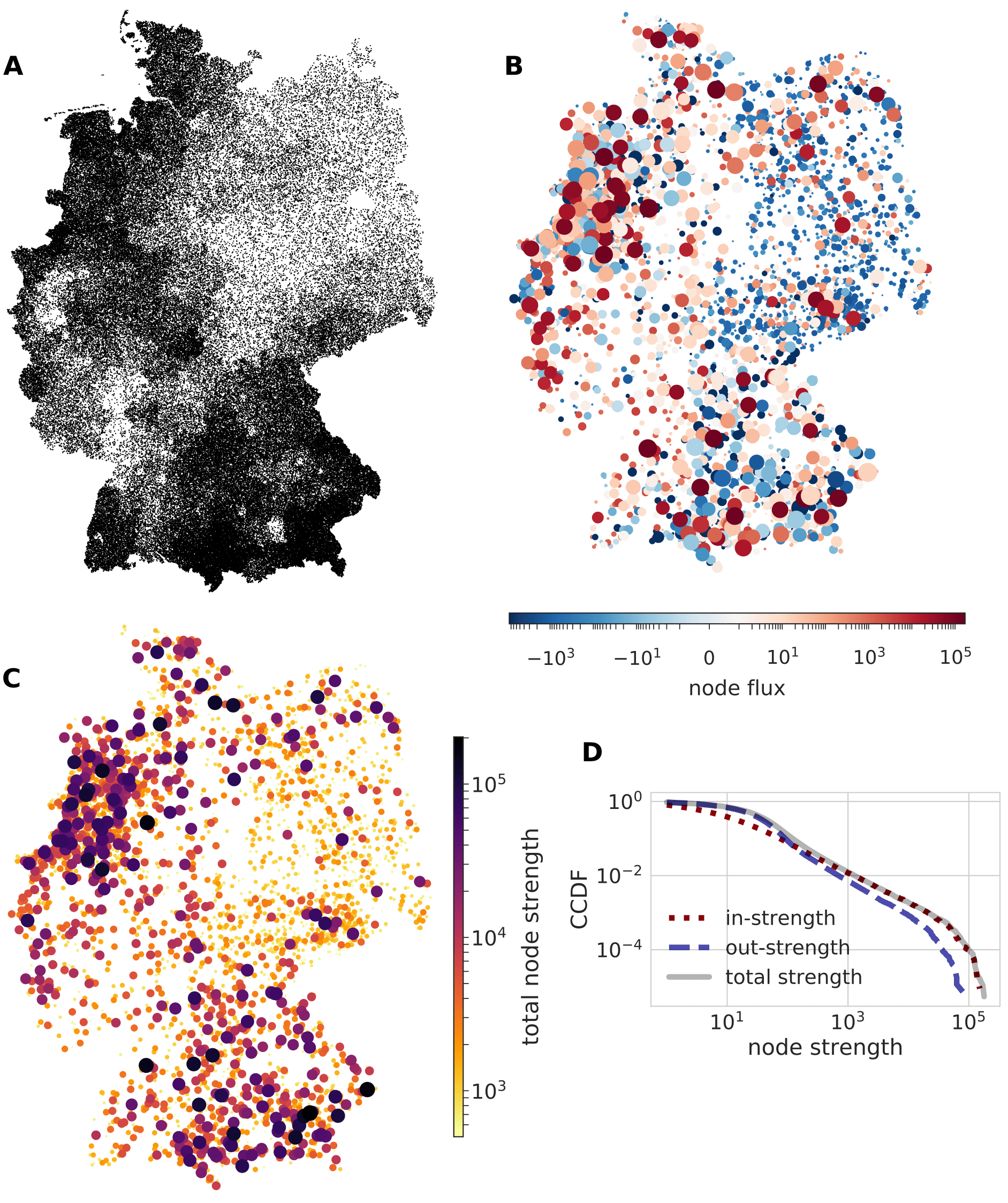}
		\caption{Node strength and flux analysis of cattle trade in the year 2010. \textbf{A}: Geographic location of all 183,454 premises that have been recorded. \textbf{B}: Color indicates net animal flux, i.e., difference between in- and out-directed animals flux. Node size corresponds to the sum of both (see subfigure C). \textbf{C}: Color and size indicate the total node strength, i.e., the sum of incoming and leaving animals. \textbf{D}: Complementary cumulative distribution function (CCDF) of the in-, out- and total strength.}
		\label{fig:network_strength_analysis}
	\end{figure}

	The geographic distribution of nodes in Fig.~\ref{fig:network_strength_analysis}A shows dense regions in the north-west and south-east including the above mentioned federal states North Rhine-Westphalia (NW), Lower Saxony (NI), Baden-W\"urttemberg (BW) and Bavaria (BY). Here, we also find the largest premises in terms of total traded animals: In Fig.~\ref{fig:network_strength_analysis}C color and size indicate the node strength, i.e., the aggregated trade volume. The heterogeneous distribution of strength becomes also apparent in Fig.~\ref{fig:network_strength_analysis}D where in-, out- and total strength is analyzed separately. Finally, we observe in Fig.~\ref{fig:network_strength_analysis}B the net flux, i.e., the difference between in- and out-directed trade volume. We display only nodes with at least 500 traded animals in Fig.~\ref{fig:network_strength_analysis}B and Fig.~\ref{fig:network_strength_analysis}C.
	
\begin{figure}[!htb]
		\centering
		\includegraphics[width=\columnwidth]{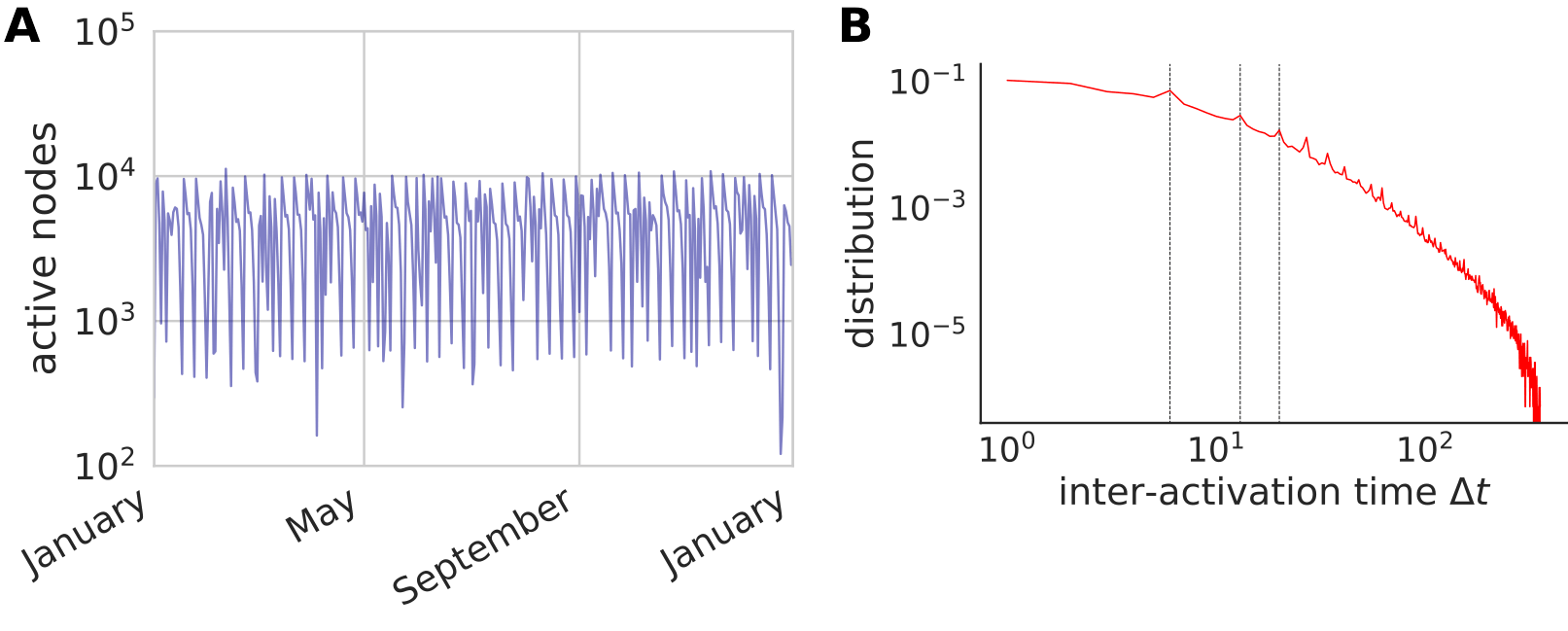}
		\caption{Temporal analysis cattle trade in the year 2010. \textbf{A} Number of premises that trade at least once on a given day. \textbf{B}: Inter-event time distribution, i.e., the interval between subsequent trade events of an arbitrary farm. Vertical lines indicate 7, 14 and 21 days.}
		\label{fig:network_temporal_analysis}
	\end{figure}

	From a temporal perspective, we find that trade fluctuates between $10^2$ and $10^4$ active nodes, i.e., farms with at least one trade event on a given day, whereas minima appear regularly on the weekends (see Fig.~\ref{fig:network_temporal_analysis}A). The weekly pattern is also apparent with a view to the inter-activation time distribution, i.e., the time interval between two successive trade events for a given node (see Fig.~\ref{fig:network_temporal_analysis}). Here, we find a broad distribution of activity with peaks around 7, 14, and 21 days.
	
	The geographic risk analysis in Sec.~\ref{sec:cattle_trade} requires us to separate the network into sub-graphs that correspond to the intra-state trade (see Fig.~\ref{fig:German_network_edges}). The largest eigenvalue of the corresponding infection propagator allows us to evaluate the outbreak risk within a federal state due to the local movement of infected animals. In Tab.~\ref{tab:bundeslaender} we list the names of all 16 federal states of Germany together with the corresponding ISO-abbreviation and basic statistics: The number of nodes, (static) edges and (temporal) contacts in the GSCC. The city states Berlin, Hamburg and Bremen as well as Saarland, which is particularly small state in terms of nodes, are marked with an Asterisk and are not considered for risk analysis in Fig.~\ref{fig:Deutschland}.
	
	\begin{table}[!htb]
		\centering
		\begin{tabular}{l|c|c|c|c}
			Federal state & ISO code & nodes & edges & contacts \\ \hline
			Schleswig-Holstein & SH & 3570 & 13541 & 49748  \\ \hline
			Hamburg$^*$ & HH & 2 & 2 & 8 \\ \hline
			Lower Saxony & NI & 12838 & 61044 & 272579 \\ \hline
			Bremen$^*$ & HB & 4 & 6 & 21 \\ \hline
			North Rhine- & & & & \\
			Westphalia & NW & 9826 & 42835 & 209677 \\ \hline
			Hesse & HE & 3622 & 12498 & 51287 \\ \hline
			Rhineland-&&&&\\
			Palatinate & RP & 1526 & 5386 & 29184 \\ \hline
			Baden-&&&&\\
			W\"urttemberg & BW & 9168 & 34434 & 150171 \\ \hline
			Bavaria & BY & 27863 & 128596 & 550047 \\ \hline
			Saarland$^*$ & SL & 26 & 52 & 343 \\ \hline
			Berlin$^*$ & BE & 0 & 0 & 0\\ \hline
			Brandenburg & BB & 715 & 2144 & 11535 \\ \hline
			Mecklenburg-&&&&\\
			Vorpommern & MV & 844 & 2852 & 21864 \\ \hline
			Saxony & SN & 690 & 1935 & 12369 \\ \hline
			Saxony-Anhalt & ST & 254 & 714 & 3422 \\ \hline
			Thuringia & TH & 344 & 957 & 5361
		\end{tabular}
		\caption{Federal states of Germany and the corresponding abbreviation together with basic network statistics: number of nodes, number of (static) edges, number of (temporal) contacts. In all cases we confined the network to the GSCC. We mark the smallest networks, i.e., the city states Berlin, Hamburg, and Bremen as well as Saarland with an asterisk.}
		\label{tab:bundeslaender}
	\end{table}
	
\begin{figure}[!htb]
		\centering
		\includegraphics[width=0.8\columnwidth]{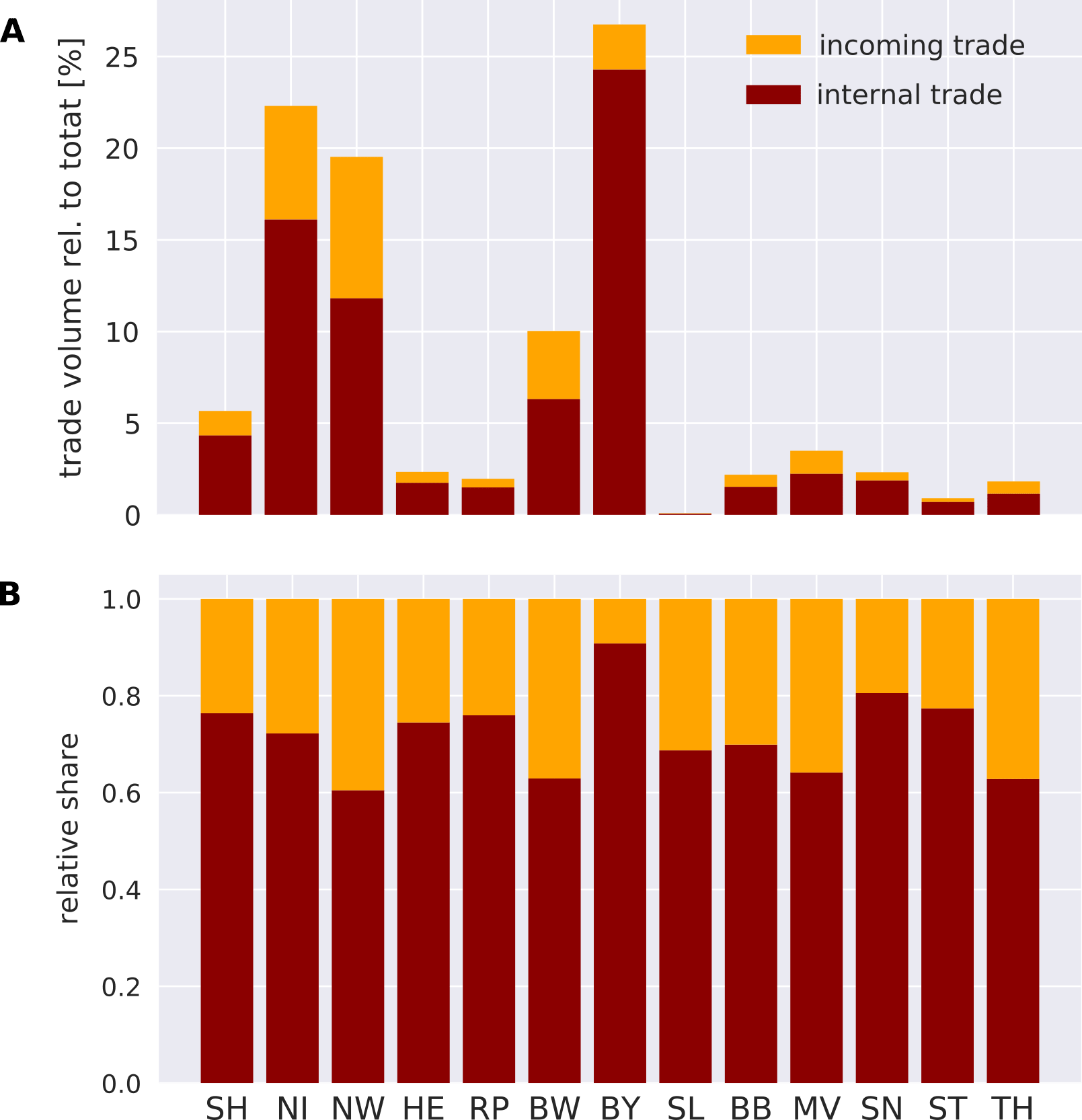}
		\caption{Error estimation after restricting trade to the federal states. \textbf{A}: Intra-state trade in terms of trade volume relative to total trade as red bars. Yellow bars on top indicate fraction of in-coming animals. All bars together sum up to $100 \%$. \textbf{B}: Sum of internal and in-coming trade volume is normalized for each federal state. See Tab.~\ref{tab:bundeslaender} for list of states and their abbreviations. \label{fig:inter_intra_state_trade}}
	\end{figure}

	Separating the trade network into subgraphs as visualized in Fig.~\ref{fig:German_network_edges}B inevitably reduces the outbreak risk as the neglected cross-border edges  would otherwise facilitate the disease transmission. In Fig.~\ref{fig:inter_intra_state_trade}A we find that a considerable fraction of trade is directed across federal states and has thus been removed. This applies in particular for the federal states NI, NW and BW. Similarly, we find that the ratio between intra-state and in-directed trade lies between 0.6 (NW) and 0.9 (BY). Thus, we conclude that a considerable fraction of trade across borders is being neglected in the geographic risk analysis in Fig.~\ref{fig:Deutschland}.
	
	It is also important to stress that we use the same parameter $\mu$ across all federal states and thus assume a uniform detection probability. In reality, federal states with a large number of premises tend to enforce stricter hygiene and intervention standards so that the actual epidemic risk for states such as Bavaria (BY) and North Rhine-Westphalia (NW) is much lower. A realistic evaluation for public health must therefore include heterogeneous recovery (detection) probabilities on the level of states or individual nodes as discussed in \cite{DAR18a} and Appendix~\ref{sec:suppl_weighted_networks} as well as a complex disease response that includes trade restrictions, increased awareness and higher bio-security.
	
	
	

	\section{Further applications} \label{sec:suppl_applications}

\begin{figure}[!htb]
		\centering
		\includegraphics[width=\columnwidth]{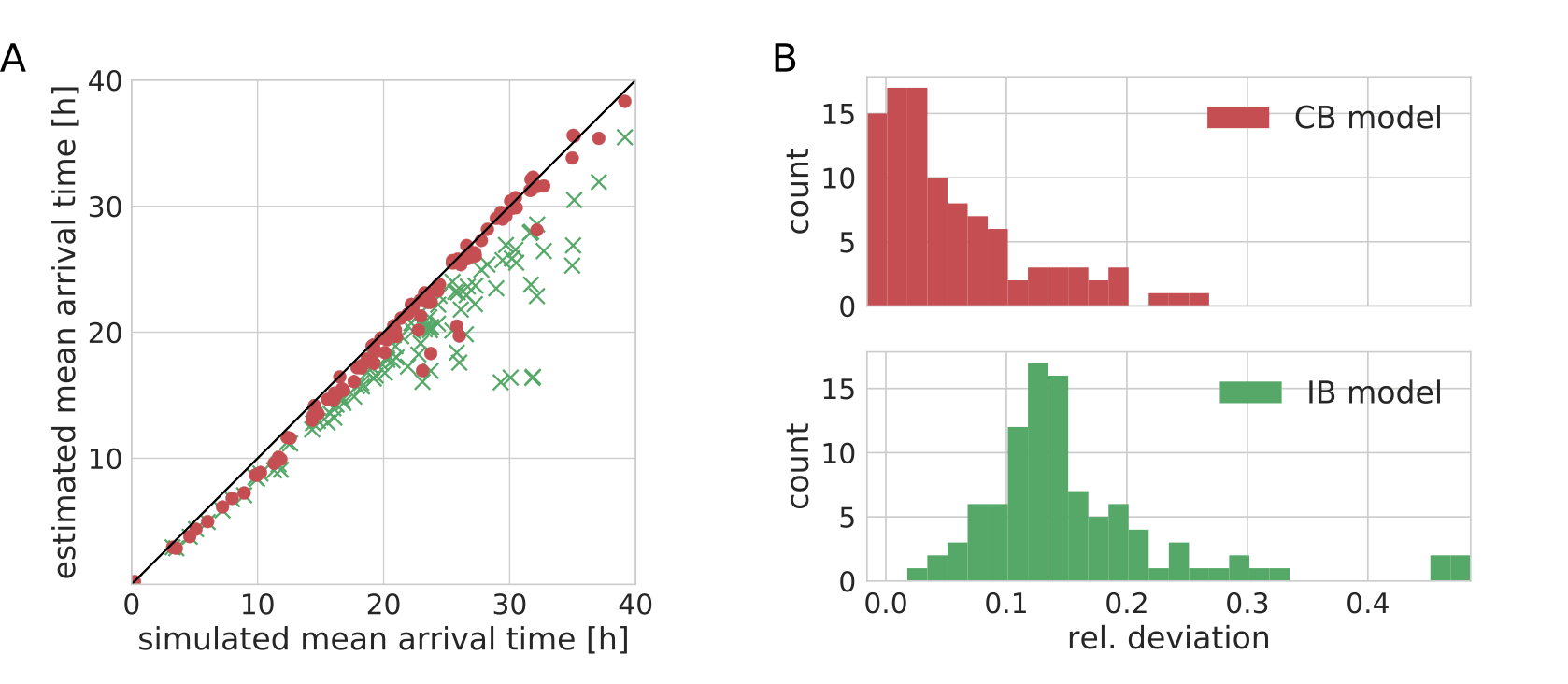}
		\caption{\textbf{A}: Comparison between simulated and estimated mean infection arrival times. We extend the data set periodically in time until the outbreak dies out. The discrete derivative of the cumulative infection probabilities (see Fig.~\ref{fig:Fig2}) yields the infection arrival probabilities of which we take the average value for every node. Results according to the CB and IB model are visualized as red circles and green crosses, respectively. The epidemic starts from the same outbreak origin and disease parameters as in Fig.~\ref{fig:Fig1}. \textbf{B}: Histogram over the relative deviation from the simulated infection arrival times. The numerical values are averaged over $10^5$ realizations.}
		\label{fig:Fig5}
	\end{figure}

    From the detailed, node-level infection trajectory we can estimate the infection arrival time from a given outbreak location to all remaining nodes. For that purpose, we extend the contact sequence periodically in time until the infection probabilities are negligible. Then, we derive the infection arrival time to a single node from the corresponding cumulative infection probability (see Fig.~\ref{fig:Fig2}) as follows: (i) The discrete derivative of the cumulative infection probability gives the probability distribution to contract the infection at a given time step. (ii) The expectation value of probability distribution gives the mean infection arrival time at the a single node, corresponding to a scatter point in Fig.~\ref{fig:Fig5}A. Here, we compare the expected values from MC simulations with the estimated infection arrival times given by the CB and IB model, respectively. In a perfect prediction the scattered values would lie on the diagonal but, as the contact network is far from a tree-like structure, the models estimate infection arrival times smaller than the observed values. The comparison between the IB and CB framework in Fig.~\ref{fig:Fig5}B shows a considerably smaller relative deviation of CB estimations from the corresponding MC simulations for the given set of disease parameters and outbreak location.
    
\begin{figure}[!htb]
		\centering
		\includegraphics[width=\columnwidth]{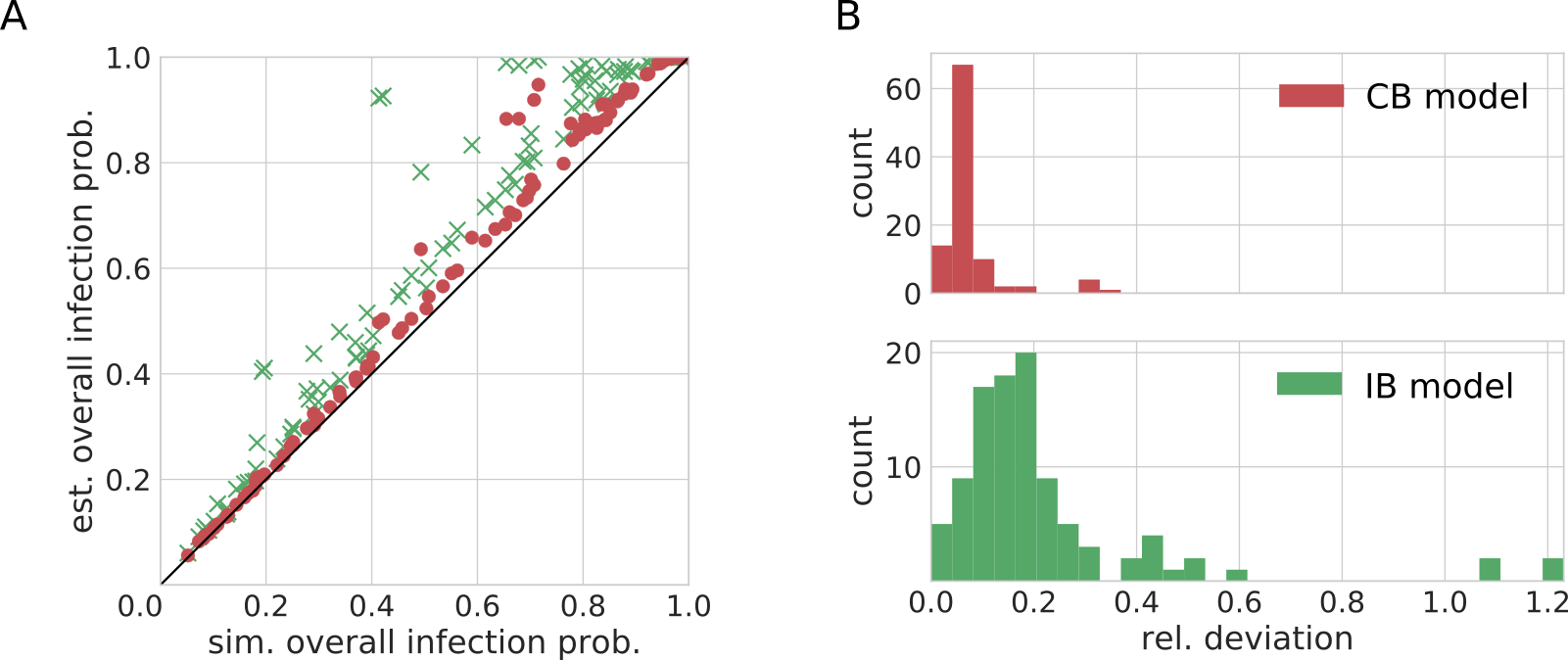}
		\caption{\textbf{A}: Comparison between simulated and estimated vulnerability. We compute the cumulative infection probability in the limit $t\rightarrow \infty$, also denoted as vulnerability. The comparison with CB and IB estimations visualized by red circles and green crosses, respectively. Each value corresponds to the vulnerability of a node given the same outbreak location and disease parameters as in Fig.~\ref{fig:Fig1}. \textbf{B}: Relative deviation of the estimated values with respect to MC simulations. The numerical values are averaged over $10^5$ realizations}
		\label{fig:Fig6}
	\end{figure}

	Another application focuses on the vulnerability of nodes with respect to a given outbreak location. Again, we assume an infinite time horizon and compare the cumulative probability that a node has been infected in the limit $t\rightarrow \infty$. As before, we find a good correlation between simulations and the estimated vulnerability in Fig.~\ref{fig:Fig6}, whereas the CB model consistently outperforms the IB approach and overestimates the expected values surprisingly little given that the underlying aggregated network is fairly dense (the average degree is $\langle k \rangle \approx 19$) and far from being tree-like. 
	
	\subsection{Trajectories averaged over outbreak locations \label{sec:averaged_outbreak_location}}
	
	For some applications, we may be interested in the trajectory of a global epidemic, averaged over outbreak locations. A sufficiently large number of initially infected nodes would then avoid complications with the early outbreak phase \cite{MIL12b, MIL14}. In this case, we adjust the MC simulations such that every node is infected independently with a given probability $1-z_l = 1 - z, ~~\forall~ l \in \mathcal{N}$ at $t = 0$. As for the analytic approach we only need to set a corresponding homogeneous initial condition and thus the computational complexity remains the same as in the previous case of one initially infected node.
	
\begin{figure}[!htb]
		\centering
		\includegraphics[width=\columnwidth]{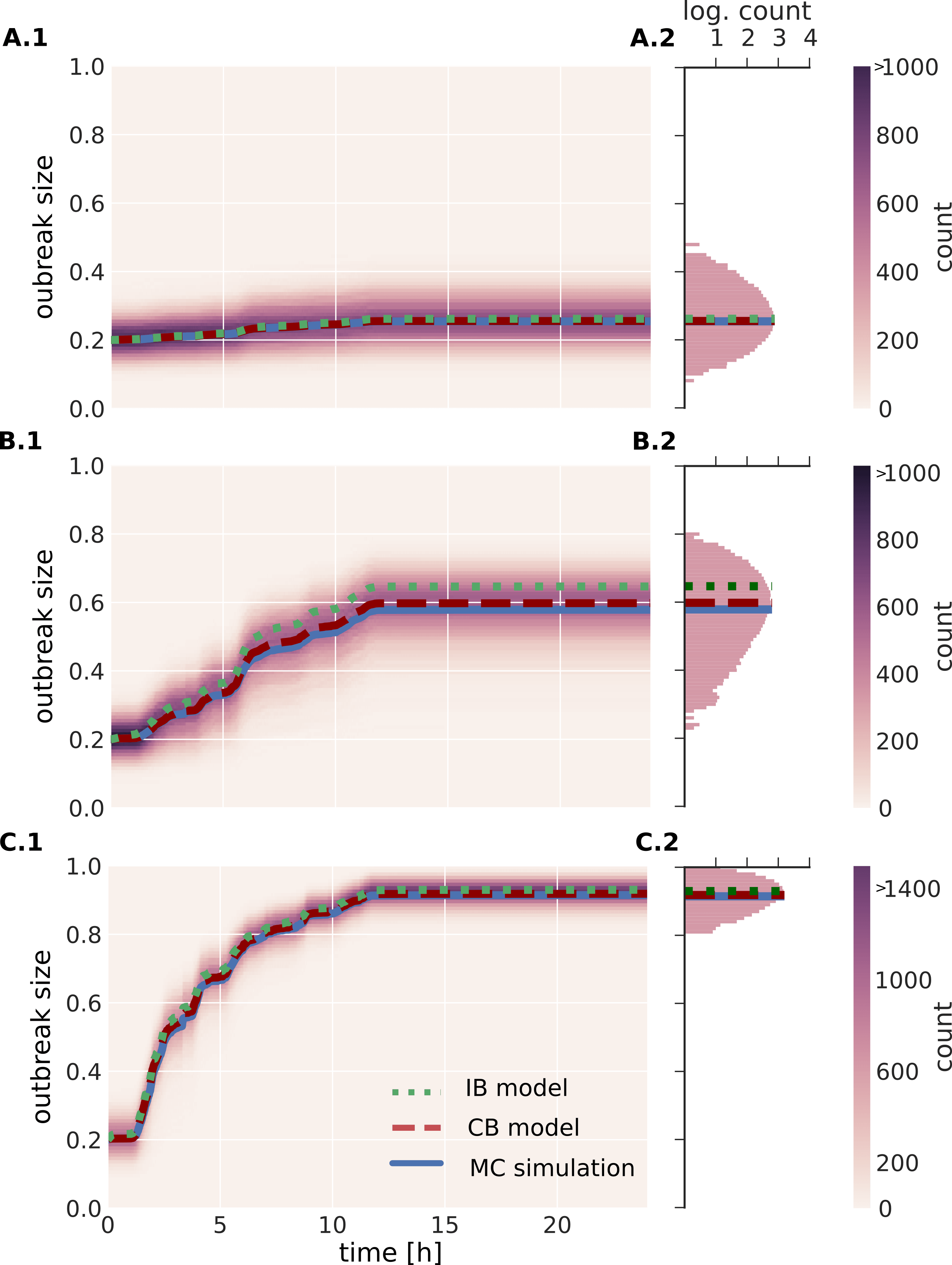}
		\caption{Cumulative infection probability with a large fraction of initially infected nodes for three different values of $\beta$ and the same outbreak location as in Fig.~\ref{fig:Fig4}A. Left column (panels A.1 - C.1): Time-evolving distribution (linear scale) of cumulatively infected individuals for infection probabilities $\beta=0.01, 0.02$, and $0.1$, respectively. We average over outbreak locations with $20\%$ of the network initially infected at random. The mean outbreak size (blue line), averaged over $10^5$ realizations and with a standard deviation below $10^{-4}$, can thus be compared to the CB and IB model (red dashed and green dotted line, respectively) with no threshold applied. Right column (panels A.2 - C.2): Final distribution (logarithmic scale) together with the averaged values.}
		\label{fig:comparison_averagedOL}
	\end{figure}

	In Fig.~\ref{fig:comparison_averagedOL} (left column) we observe a narrow, time-dependent distribution of cumulatively infected nodes around the mean value for three different infection probabilities. Without applying any additional threshold, we find a close agreement between the averaged trajectory, the CB and IB model in all cases. In contrast to Fig.~\ref{fig:Fig4} of the main text, we observe in Fig.~\ref{fig:comparison_averagedOL} (right column) only one peak in the distribution, due to the large number of initially infected nodes.
	
\begin{figure}[!htb]
		\centering
		\includegraphics[width=\columnwidth]{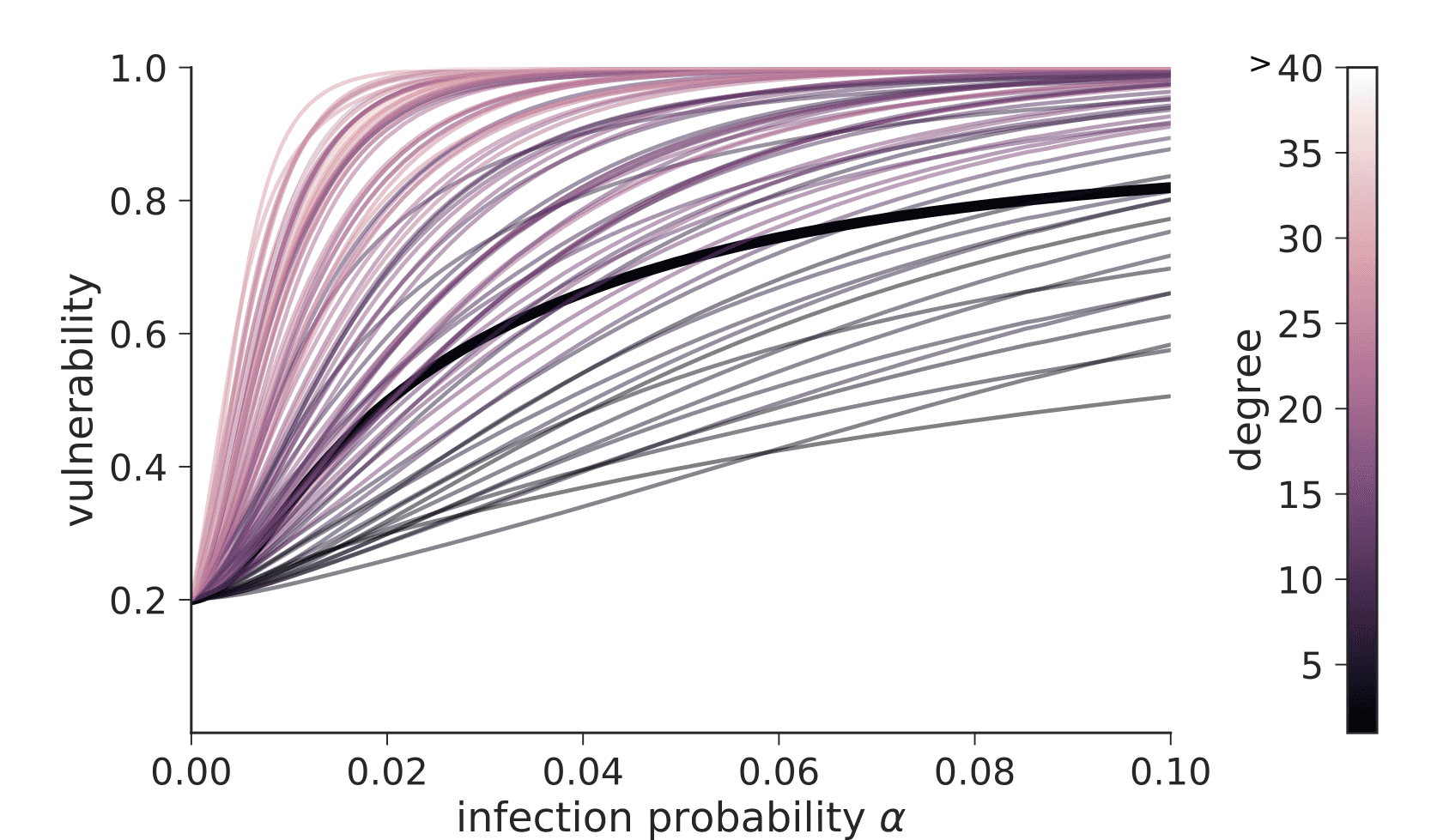}
		\caption{Vulnerability as a function of the infection probability $\beta$ estimated from the CB model. Each curve represents the vulnerability of a node, i.e., the probability to contract the infection from a set of randomly chosen outbreak locations. Here, we estimate the vulnerability according to the CB model. Starting from an initial infection probability of $1-z=0.2$ we propagate the infection over time until convergence. We stop when the largest increase in vulnerability after 24h falls below $10^{-3}$. The colors indicate the degree of each node in the underlying time-aggregated graph. Moreover, the behavior of one selected node is highlighted.}
		\label{fig:vulnerability_curves}
	\end{figure}

	One potential application is to calculate the vulnerability of a node as discussed in \cite{ROG15}. Here, the vulnerability is defined as the probability that a given node is eventually infected by a disease that started somewhere in the network. The value can be used to rank nodes in order to prioritize surveillance or vaccination measures to the nodes that are most likely to contract the disease when resources are limited. In Fig.~\ref{fig:vulnerability_curves} every curve represents the vulnerability of one node as a function of the infection probability $\beta$. These results are derived from the CB model, given an initial infection probability of $1-z=0.2$.  The individual colors correspond to the degree of the node in the underlying time-aggregated graph and serve as a guide to the eye. Interestingly, we find that the ranking, as estimated by the CB model, may change with increasing infection probabilities $\beta$ as can be seen from the highlighted curve in Fig.~\ref{fig:vulnerability_curves}. This effect has been observed earlier in the context of static networks \cite{ROG15} and indicates that network properties alone are often not sufficient to rank nodes as they do not take into account details of the dynamic system.
	
	\subsection{Additional numerical results}
	
\begin{figure*}[!htb]
		\centering
		\includegraphics[width=\textwidth]{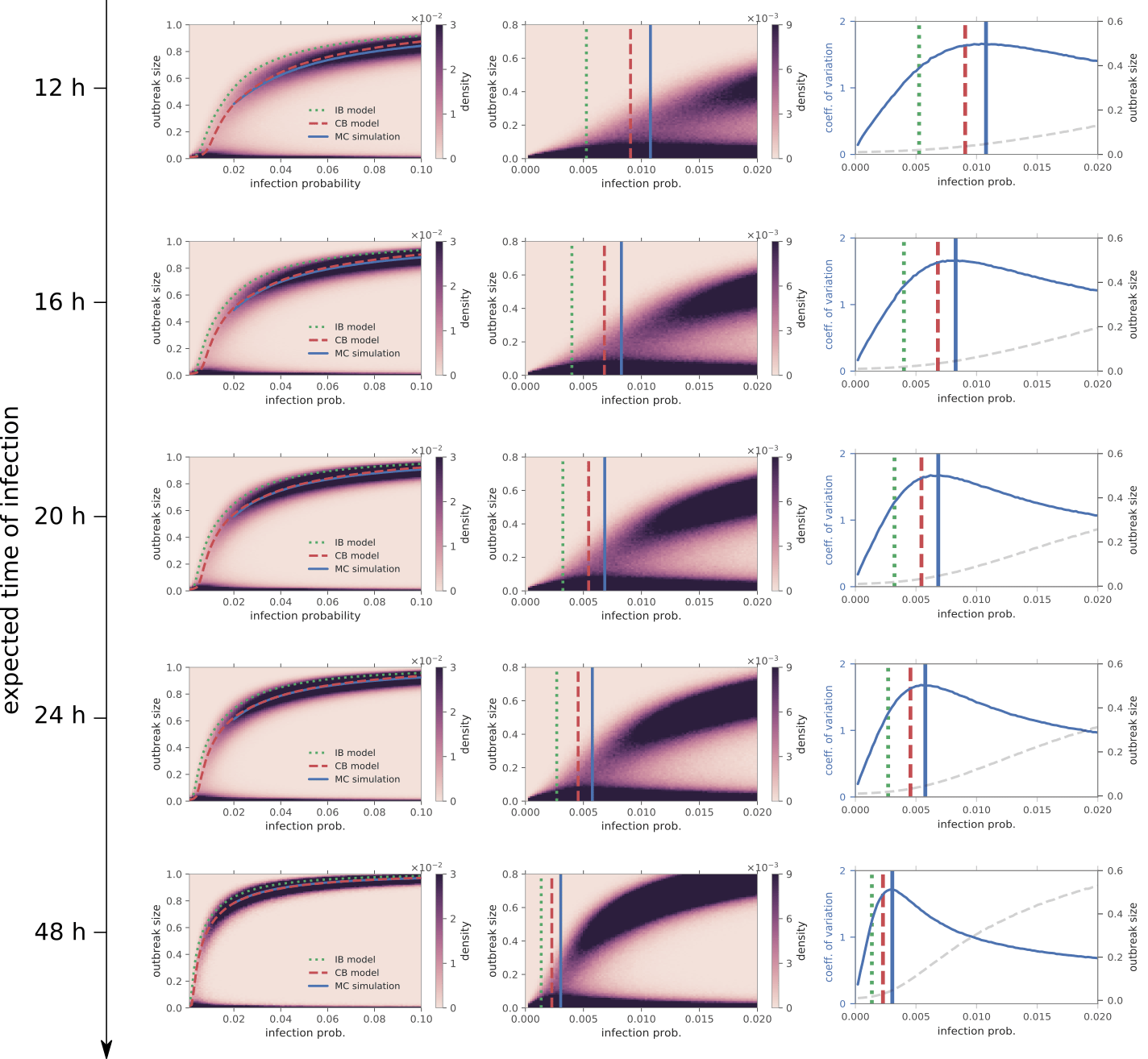}
		\caption{Comparison between MC simulations and the mean-field models for different values of $\mu$. Every row assumes a fixed recovery probability $\mu$ with decreasing values from top to bottom: $\mu = 4.63 \cdot 10^{-4}$, $\mu = 3.47 \cdot 10^{-4}$, $\mu = 2.78 \cdot 10^{-4}$, $\mu = 2.31 \cdot 10^{-4}$, and $\mu = 1.16 \cdot 10^{-4}$. These values correspond to an expected infection period of 12, 16, 20, 24, and 48 hours. Left column, middle column and right column correspond to Fig.~\ref{fig:attack_rates_HT09}, Fig.~\ref{fig:susceptibility_HT09}A, and Fig.~\ref{fig:susceptibility_HT09}B, respectively, of the main text.}
		\label{fig:sociopatterns_comparison}
	\end{figure*}

	The analysis of the conference data set in the main text was limited to a single value of the recovery probability with $\mu = 2.85 \cdot 10^{-4}$. This choice corresponds to an expected infectious period of about $19.5$ hours. In addition to the analysis of the main text, we present in Fig.~\ref{fig:sociopatterns_comparison} similar results for different values of $\mu$. The left, middle and right column correspond to Fig.~\ref{fig:attack_rates_HT09}, Fig.~\ref{fig:susceptibility_HT09}A, and Fig.~\ref{fig:susceptibility_HT09}B, respectively. In all cases, the CB model gives a closer bound to MC simulations as compared to the IB approach.
	
\begin{figure*}[!htb]
		\centering
		\includegraphics[width=\textwidth]{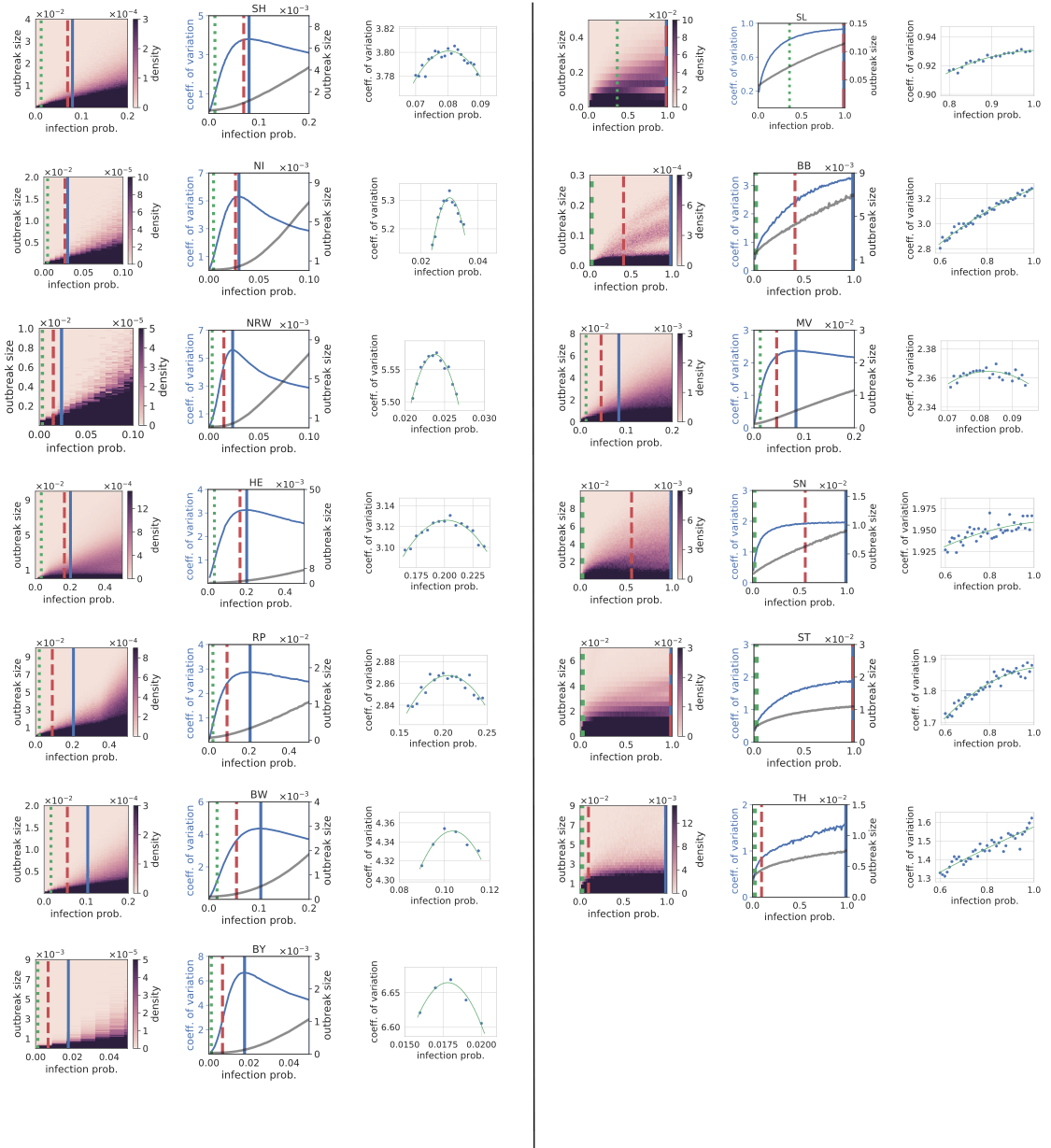}
		\caption{Detailed analysis of the epidemic threshold for the cattle trade network with $\mu = 1/28$. The central black line separates the figure into two panels. Every row (in a panel) provides results for a single federal state in Germany, excluding the city states Berlin, Bremen, and Hamburg. Left column: outbreak size distribution as a function of the infection probability $\beta$. Vertical blue, red and green lines mark the critical infection probability according to MC simulations, the CB, and IB model, respectively. Middle column: Coefficient of variation $c_v$ (blue line, left axis), mean outbreak size (gray dashed line, right axis). Right column: Selected values of $c_v$ for infection probabilities close to the critical value. Quadratic fit (green line) estimates the maximum and hence $\beta_{\text{crit.}}$.}
		\label{fig:HIT_threshold_analysis_28days}
	\end{figure*}

	Next, we complete the analysis in Fig.~\ref{fig:cattle_threshold_comparison} of the main text on the epidemic threshold for the example of the animal trade network. Again, we assume $\mu = 1/28$ and provide in Fig.~\ref{fig:HIT_threshold_analysis_28days} a detailed analysis of all federal states, excluding the city states Berlin, Hamburg, and Bremen. The left and middle column in Fig.~\ref{fig:HIT_threshold_analysis_28days}, provide a similar analysis to Fig.~\ref{fig:susceptibility_HT09}A, and Fig.~\ref{fig:susceptibility_HT09}B of the main text, respectively. In other words, we present the distribution of outbreak sizes for different values of the infection probability $\beta$ (left column) from which we derive the coefficient of variation $c_v$ (blue line, middle column). The right column presents values of $c_v$ that are close to the peak and a quadratic fit (green line, right column) that determines the numerical estimation of the critical infection probability (blue vertical line). This value can be compared to spectral estimations from the mean-field models. In agreement with previous results, we find that the criticality condition in Eq.~\ref{eq:threshold} of the CB model (see main text) improves previous results of the IB approach.
	
\begin{figure*}[!htb]
		\centering
		\includegraphics[width=\textwidth]{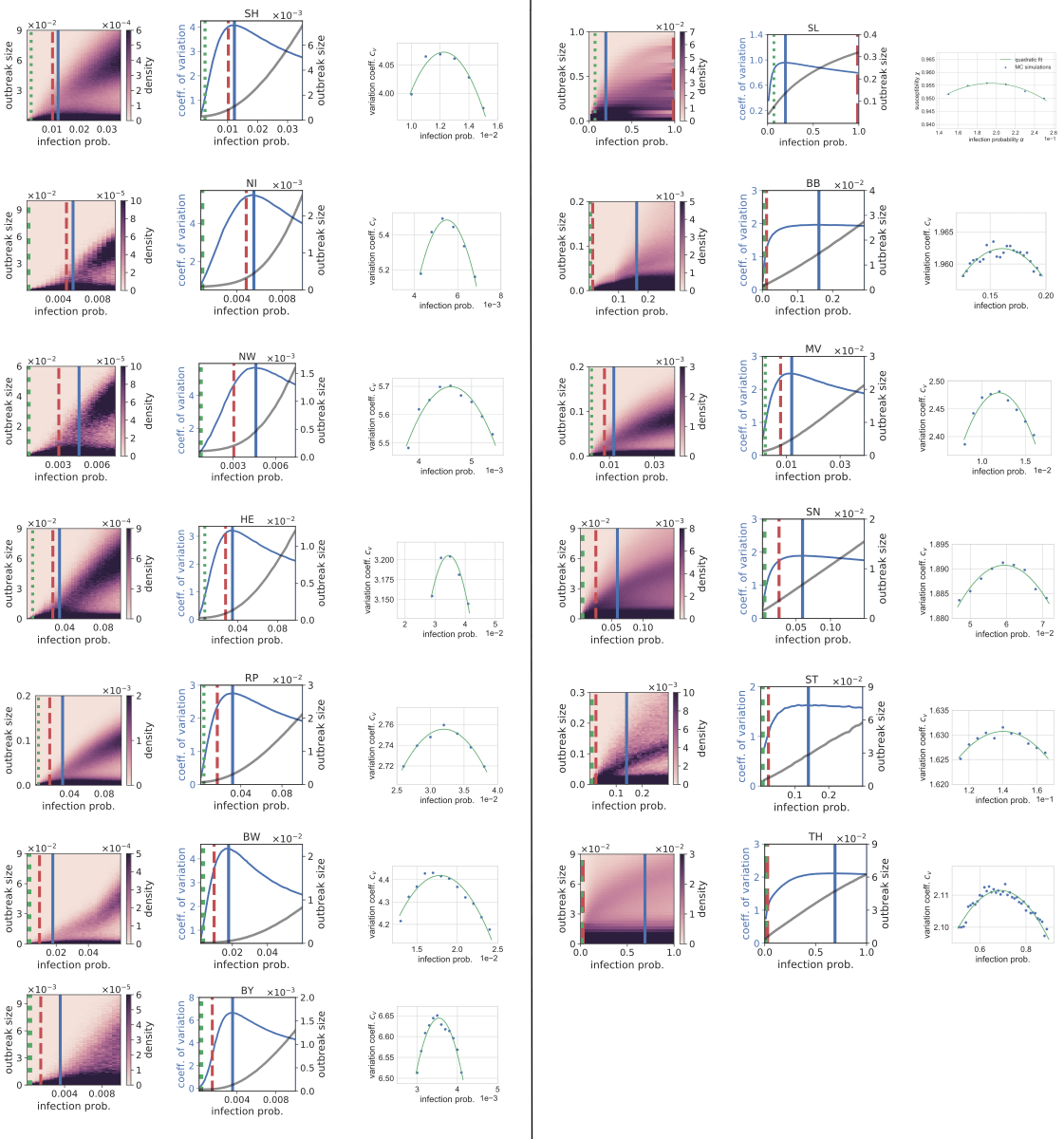}
		\caption{Detailed analysis of the epidemic threshold for the cattle trade network with $\mu = 1/120$. The analysis is akin to Fig.~\ref{fig:HIT_threshold_analysis_28days}.}
		\label{fig:HIT_threshold_analysis_120days}
	\end{figure*}

	Finally, we provide an additional analysis of  the epidemic threshold for the cattle trade data with $\mu = 1 / 120$. Results in Fig.~\ref{fig:HIT_threshold_analysis_120days} are akin to our previous analysis in Fig.~\ref{fig:HIT_threshold_analysis_28days}, except for Saarland (SL). Here, the spectral condition in Eq.~\ref{eq:threshold} of the CB model predicts that every outbreak remains localized, i.e. $\beta^{\text{CB}}_{\text{crit.}} = 1$, whereas MC simulations suggest a transition to global epidemics, hence $\beta^{\text{MC}}_{\text{crit.}} < 1$. We attribute the inconsistency to the small size of the network (26 nodes). The spectral approach assumes implicitly an infinitely large network, which is clearly violated in this case.
	
\begin{figure}[!htb]
		\centering
		\includegraphics[width=\columnwidth]{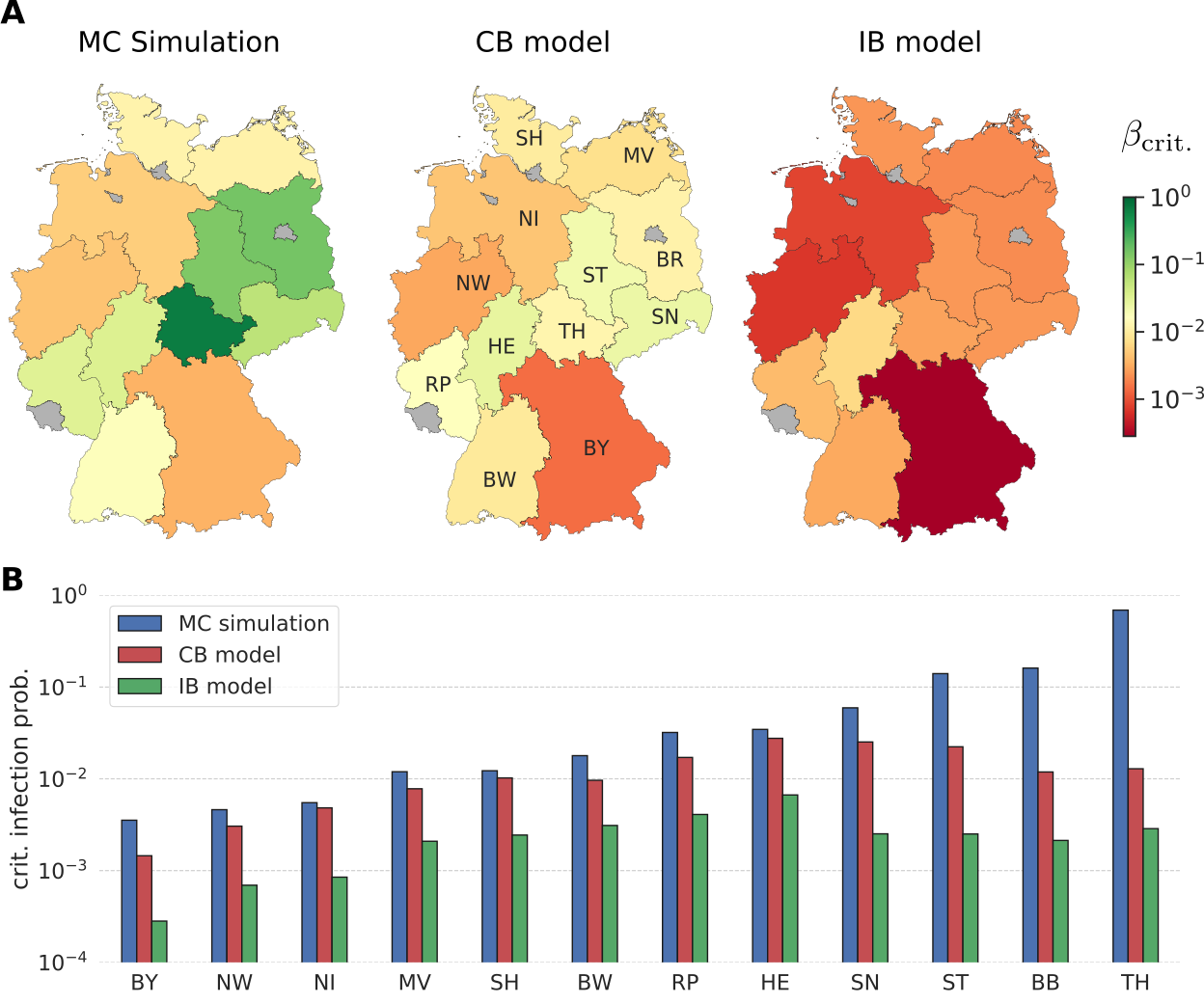}
		\caption{\textbf{A}: Spatial variation of epidemic risk for $\mu = 1 / 120$. The critical infection probability $\beta_{\text{crit}}$ determines the color of the federal states (see Fig.~\ref{fig:HIT_threshold_analysis_120days} for details). The visualization is akin to Fig.~\ref{fig:Deutschland} of the main text.}
		\label{fig:Deutschland_120days}
	\end{figure}

	We summarize the results for $\mu = 1 / 120$ in Fig.~\ref{fig:Deutschland_120days}, akin to Fig.~\ref{fig:Deutschland} of the main text. The risk map in Fig.~\ref{fig:Deutschland_120days}A visualizes the spatial variability of the outbreak risk and each group of blue, red and green bars in Fig.~\ref{fig:Deutschland_120days}B provide a quantitative comparison between MC results, the CB model and the IB approach, respectively.

	\end{document}